# Experimentally validated and empirically compared machine learning approach for predicting yield strength of additively manufactured multi-principal element alloys from Co-Cr-Fe-Mn-Ni system

Abhinav Chandraker[1], Sampad Barik[1], Nichenametla Jai Sai[2], Ankur Chauhan[1]*

[1]Extreme Environments Materials Group, Department of Materials Engineering, Indian Institute of Science (IISc), Bengaluru, 560012 Karnataka, India

[2]Department of Materials Science and Metallurgy, University of Cambridge, 27 Charles Babbage Road, Cambridge, CB3 0FS, UK

*Corresponding author: ankurchauhan@iisc.ac.in

## Abstract

Traditionally, yield strength prediction relies on detailed and resource-intensive microstructural characterization combined with empirical equations. However, quantifying microstructural feature length scales for novel processes like additive manufacturing, which involves inhomogeneous hierarchical features, poses a challenge. The lack of accurate material constants for broader composition ranges further limits empirical predictions. This study proposes an alternative machine learning (ML) approach for predicting the yield strength of additively manufactured (AM) multi-principal element alloys (MPEAs) from the Co-Cr-Fe-Mn-Ni system by correlating composition, printing parameters, and testing conditions. The best-performing ML model achieved an $R^2$ of 0.84, comparable to that achieved using microstructural detail-driven empirical strengthening contributions. The validity of the ML approach was further confirmed by printing and testing two compositions (one novel and one from the dataset). This data-driven approach directly relates yield strength to initial printing parameters, highlighting their significance and individual effects, such as scan velocity's direct impact and laser power's inverse impact on yield strength. This demonstrates ML's potential to guide AM processes, reducing the need for iterative experiments and enabling rapid exploration of compositional and printing spaces to achieve desired properties.

## 1. Introduction

In recent years, additive manufacturing has emerged as a promising technique, offering advantages such as producing complex near-net shape products in a single step. However, additive manufacturing also presents challenges. The quality and characteristics of the printed material are heavily influenced by specific printing conditions[1]. Additionally, the printing technique significantly affects the microstructure and properties of the printed material[2,3]. For example, the laser melt deposition (LMD) technique typically creates deeper melt pools due to slower scanning speeds compared to laser powder bed fusion (LPBF), leading to variations in texture, dislocation cell sizes, chemical heterogeneity, and melt pool geometry[4]. These factors introduce considerable variability in the resulting mechanical properties. Controlling properties like yield strength (YS) is crucial, especially considering the stringent standards required for industrial applications. Therefore, accurate prediction of YS is essential, which will also aid in rapidly exploring the compositional and printing parameter space to obtain prints with improved properties.

Traditional methods for predicting YS rely on empirical relations, incorporating contributions from various strengthening sources such as solid solution, dislocation forest, grain/twin boundaries, and precipitates/dispersoids. These empirical predictions require extensive microstructural characterization, which is labour-intensive and destructive. Developed initially for conventionally manufactured alloys with relatively homogeneous microstructures that are easier to characterize, these estimations face challenges when applied to additively manufactured (AM) parts. AM parts exhibit a hierarchical microstructure characterized by irregular and textured grains and a dense network of sub-grain boundaries, often exhibiting elemental segregation. Consequently, estimating boundary strength contributions using the Hall-Petch equation becomes challenging due to difficulties in accurately quantifying grain sizes in AM parts, which feature columnar grains of varying sizes. Moreover, these empirical estimations rely on precise materials-specific constants (e.g., shear modulus, lattice friction, Hall-Petch constant, Burgers vector, Taylor factor), which are currently unavailable for novel compositions or have not been thoroughly measured. As a result, approximations from similar compositions are often used, potentially impacting the accuracy of empirical predictions. Additionally, the reported values for material constants, such as lattice friction and Hall-Petch constants, exhibit significant variation in the literature[5][6], highlighting challenges in their accurate determination[5][6]. Moreover, this approach cannot direct the manufacturing process by correlating the initial process parameters with material properties. Given these challenges and constraints, a data-driven prediction approach incorporating the non-linear and higher-order effects of initial parameters may prove more effective.

Machine learning (ML) excels at this and has recently become popular for designing and optimizing additive manufacturing prints[7–9]. While studies have focused on predicting the mechanical properties of AM materials[10,11], few have achieved this without relying on detailed microstructural information. Moreover, existing research primarily revolves around traditional alloys, such as stainless steels[12,13] and Ti alloys[14], which have a limited compositional range. Only recently have studies focused on employing

ML to predict the mechanical properties of novel multiple-principal element alloys (MPEAs), which have a broad compositional space capable of achieving various strength and ductility combinations. However, these studies considered MPEAs fabricated through conventional routes[15–18][19]. For instance, Wen et al.[20] used an iterative strategy combining ML and experimental feedback to search for novel alloys with high hardness in the AlCoCrCuFeNi system. Yang et al.[21] have also carried out the hardness prediction of several MPEAs using a support vector machine (SVM) algorithm, employing several compositional descriptors and feature selection strategies. Furthermore, Sai et al.[22] predicted fatigue lives of Co-Cr-Fe-Mn-Ni and Al-Co-Cr-Fe-Mn-Ni systems using random forest (RF), support vector regression (SVR), gradient boosting (GBoost), and extreme gradient boosting (XGBoost) algorithms.

Despite these efforts, there is a scarcity of research focusing on modeling the YS of MPEAs fabricated through the additive manufacturing route[13,23], which will enable rapid screening of the desired printing parameters and compositions from a vast space, significantly reducing the number of experimental iterations required. Therefore, this study utilizes ML to predict the YS of the MPEAs from the Co-Cr-Fe-Mn-Ni system manufactured via two additive manufacturing techniques: LMD and LPBF. Various alternate terminologies, such as laser engineering net shape (LENS) and directed energy deposition (DED), have also been grouped under LMD. This approach incorporated compositional, printing, and testing parameters as input features, filling a gap identified in previous research[19]. The best model's effectiveness was experimentally validated and compared with empirical predictions obtained by accounting for all potential strengthening contributions. Additional insights were gained into the significance and individual effects of the input features, and a synthetic dataset was used to analyze the impact of specific printing parameters on the YS trends. This ML framework proposes a pathway for guiding future experiments to select optimal compositions and printing parameters to achieve customized properties.

## 2. ML methodology

Figure 1 illustrates an overview of the methodology utilized in this study. It involved compiling a dataset of the AM MPEAs from the Co-Cr-Fe-Mn-Ni system from published literature[24–98]. Ensemble-based ML algorithms such as RF, GBoost, XGBoost, and SVR were employed and optimized to achieve robust and accurate predictions. Two alloys were printed using the LMD technique and subjected to tensile testing to experimentally validate the best-performing ML model. The ML model predictions were also compared with empirical predictions obtained using microstructural detail-driven strengthening contributions. To gain insights into the importance of the input features on the YS and their likely roles for LPBF and LMD printing routes, the SHapley Additive exPlanations (SHAP) algorithm was utilized. Additionally, a synthetic dataset was employed to analyze the impact of individual printing parameters on the YS trends.

Detailed descriptions of all algorithms used can be found in the supplementary material section 6. The scikit-learn package[99] in Python was utilized to implement the modelling process.

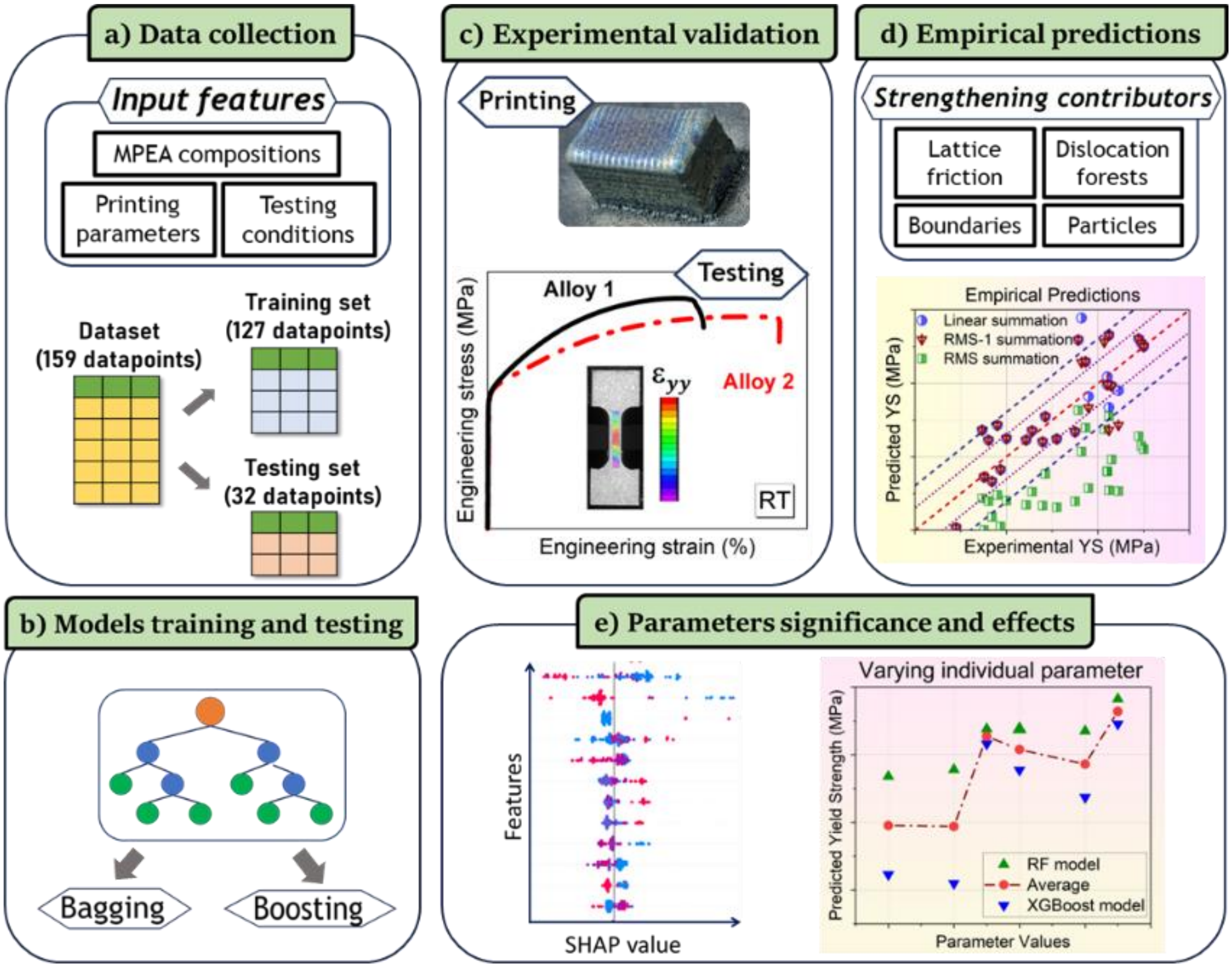

Figure 1: Overview of the methodology, comprising five steps: a) Dataset collection from the published literature, b) Development of ML models using training and testing datasets, c) printing and testing two alloys for experimental validation, d) empirical predictions of YS by considering various strengthening mechanisms contributions and comparison with ML predictions, and e) investigating parameters significance and effect through SHAP analysis and construction of feature trends.

## 3. Experimental procedure

$Co_{33.3}Cr_{33.3}Ni_{33.3}$ and $Co_{25}Cr_{45}Ni_{30}$ alloys were fabricated using the LMD technique. Commercially pure elemental powders (with ~ 99.6% purity) of spherical morphology and having a diameter in the range of 50 to 100 μm were utilized. The powders were mechanically mixed in the required proportions using a planetary ball mill with a 5:1 ball-to-powder ratio at 100 rpm for 3 hours in an argon gas atmosphere. The mixed powders were then fed in a FormAlloy 3D printer to create blocks with dimensions of 26 mm length, 14 mm width, and 12 mm height, using a laser beam diameter of 1 mm, powder feed rate of 7.54 g/sec, laser power of 700 W, scan velocity of 8 mm/sec, hatch spacing of 0.3 mm and a 67° rotation between the layers.

A Bruker's D8 Advance X-ray diffractometer with copper source (Cu-Kα radiation $\lambda = 0.154$ nm) was employed for the phase analysis. X-ray diffraction (XRD) patterns were acquired using the following parameters: 2θ-range from 30° to 100°, step size $\Delta 2\theta = 0.02°$. Flat dog-bone-shaped specimens with a gauge length of 6 mm, gauge width of 2 mm, and thickness of 2mm were cut using electrical discharge machining. For microstructure characterization, all specimens' gauge sections were ground up to P4000 SiC emery paper, followed by cloth polishing using colloidal silica, and finally etched with a Kalling's 2 Reagent consisting of 40 ml ethyl alcohol, 40 ml hydrochloric acid (35-38%), and 2 gm copper (II) chloride dihydrate. Optical micrographs were obtained using a Leica DMi8 microscope. Tensile tests were conducted on an Instron universal testing machine (UTM) at 298 K with a strain rate of $10^{-3}$ $s^{-1}$. Strain measurements were taken using Digital Image Correlation (DIC) with Vic2D software. Tensile sample faces were sprayed with white and black paint to create a speckle pattern for DIC. At least two tests were conducted per condition to ensure statistical accuracy.

## 4. ML execution

### 4.1. Data compilation and processing

Data from studies reporting dense prints were selected to eliminate porosity effects. The dataset comprised 159 data points, with 83 of LPBF and the rest of LMD. It covered composition features, printing parameters, testing conditions, and tensile YS. Some of the missing feature values were synthetically fed, the details of which are provided in the supplementary material section 1. Microstructure-based input features were bypassed to avoid needing experimental characterization for prediction. Additionally, such details were inconsistently reported in the literature.

The selection of descriptors significantly impacts the performance of the ML model. Ward et al.[100] proposed frameworks for creating compositional features to predict the properties of inorganic materials. This study utilized a broad range of elemental and functional descriptors to effectively capture the physical and chemical influences of composition, lattice arrangement, and atomic bonding. Elemental descriptors such as atomic size ratio, ionization energies, number of valence electrons, etc., and functional descriptors such as local size mismatch, Peierls Nabarro factor, lattice distortion energy, etc., were considered. Details of the compositional descriptors utilized were obtained from previous works[19–21,101,102] and are provided in the supplementary material section 3.

Printing parameters included laser power, input energy density, scan pattern rotation, and scan velocity. Given the examination of two printing techniques, separate input energy density features were employed: volumetric energy density (VED) for the LPBF technique and linear energy density (LED) for the LMD technique, defined by equations 1 and 2, respectively.

$$VED = \frac{P}{v * d * t} \quad \text{Equation 1}$$

where $P$ stands for laser power (W), $v$ stands for scan velocity (mm/s), $d$ stands for scan spacing (µm), and $t$ stands for the layer height of the track (µm).

$$LED = \frac{P}{v * D} \quad \text{Equation 2}$$

where $D$ is the laser beam diameter (µm).

The input features for testing conditions include the specimen's cross-section, test temperature, strain rate, and the testing direction relative to the printing direction. These features provide context for the conditions under which the MPEAs were tested. The target output feature for the prediction is the YS of the as-printed MPEAs, which ranged from 70 to 872 MPa. Categorical features like scan pattern rotation and testing direction were encoded with numerical values to facilitate their use by ML algorithms. Table 1 summarizes the numerical values of the printing and testing input features, along with their respective minimum and maximum values. Table 2 details the specific numerical assignments for the categorical features. Analysis of feature distribution revealed sparse data for certain features, such as strain rate and tensile testing directions (see supplementary material section 2).

Table 1: Numerical values of the printing and testing input features and their respective minimum and maximum values.

| Numerical input feature (unit) | Minimum value | Maximum value |
|---|---|---|
| **Laser power (W)** | 50 | 1625 |
| **Scan velocity (mm/sec)** | 3 | 2500 |
| **LPBF VED (J/mm$^3$)** | 44 | 225 |
| **LMD LED (J/mm$^2$)** | 17 | 223 |
| **Testing strain rate (s$^{-1}$)** | 0.00001 | 0.06667 |
| **Testing temperature (°C)** | -204 | 800 |
| **Cross section area (mm$^2$)** | 0.8 | 50.24 |

Table 2: Categorical features and list of their corresponding values.

| Categorical feature | Values |
|---|---|
| **Scan pattern rotation** | 0°, 45°, 67°, 90°, 180°, 45°-checkboard, 67°-checkboard, 90°-checkboard |
| **Tensile test direction (with respect to printed material)** | X without scan pattern rotation, Y without scan pattern rotation, X with scan pattern rotation, 45° to X and Y, Z, 45° to X and Z |

## 4.2. Model training methodology

To avoid bias from differing magnitudes, the data was standardized (see supplementary materials section 2) before splitting into 80% training and 20% test datasets. Four ensemble algorithms—RF, GBoost, XGBoost, and SVR—were utilized to build the ML models. The algorithms were trained on the training dataset, and hyperparameter tuning was performed using a Bayesian search algorithm on the training dataset to construct optimized models. This advanced technique systematically evaluates multiple sets of hyperparameters sequentially in an informed manner, identifying the hyperparameters that best capture the dataset characteristics. The optimized hyperparameters for each model are provided in the supplementary material section 8. Model performance was assessed using two metrics: Coefficient of determination ($R^2$) and Root Mean Square Error (RMSE), defined by Equations 3 and 4, respectively.

$$R^2 = 1 - \frac{\sum(YS_{experimental} - YS_{predicted})^2}{\sum(YS_{experimental} - YS_{mean})^2} \qquad \text{Equation 3}$$

$$RMSE = \sqrt{\frac{\sum(YS_{experimental} - YS_{predicted})^2}{Z}} \qquad \text{Equation 4}$$

where $YS_{experimental}$ is the experimental YS, $YS_{predicted}$ is the predicted YS, $YS_{mean}$ is the mean experimental YS, and $Z$ is the count of experiments.

These metrics provide insights into prediction accuracy. A high $R^2$ value indicates the model captures actual data variations well, while a low RMSE value signifies minimal deviation between predicted and actual values.

Although including all input features is important for building a robust model, we limited the number of input features due to the dataset size. Introducing more features could expand the input space dimensionality, potentially leading to the "curse of dimensionality," which can cause overfitting and reduce model performance[103]. Therefore, feature selection was performed to identify the key input features influencing the output variable. This was carried out using only the training dataset to prevent the test dataset from affecting the modeling process, ensuring the integrity of the model's performance assessment. More details on input feature selection are provided in the supplementary material section 5. The model with 16 input features showed the best performance and is, therefore, considered for further investigations.

# 5. Model performance assessment

All models except SVR were well-trained using the 16 selected input features, achieving train $R^2$ values above 0.85 and train RMSEs below 75 MPa, as shown in Figure 2. RF performed the best on the test dataset, with a test $R^2$ of 0.84 and a test RMSE of 61.0 MPa. XGBoost followed with a test $R^2$ of 0.74 and test RMSE of 77.3 MPa.

Given the sparsity of the dataset, it was crucial to evaluate the robustness of the optimized model. Consequently, the sensitivity of model performance to data split choice was also investigated (see supplementary materials section 9). The mean test $R^2$ remained above 0.70 for the RF and XGBoost models across 25 additional data splits, suggesting low data split sensitivity.

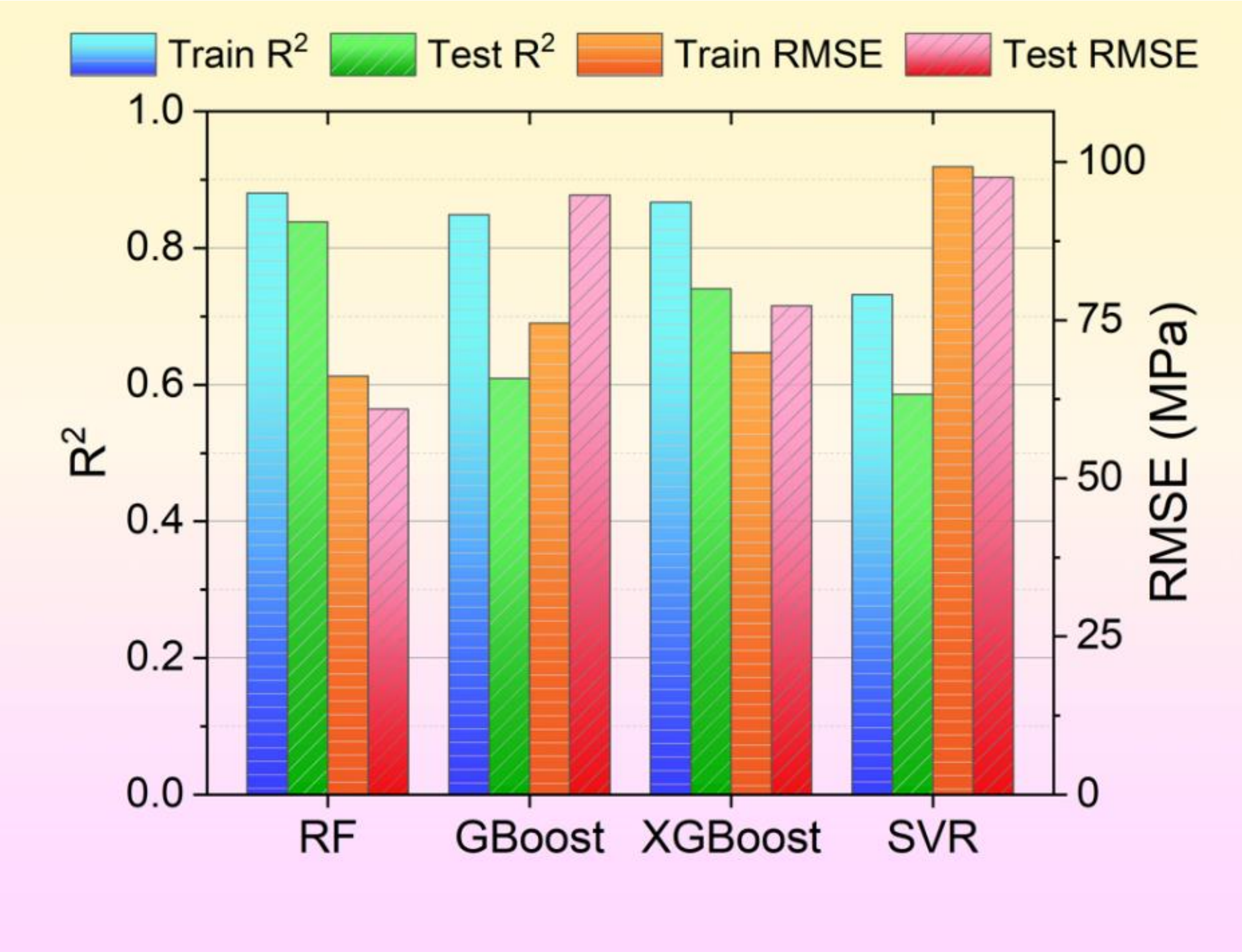

Figure 2: Coefficient of determination (R2) and Root Mean Square Error (RMSE) values for the employed models. All models except SVR are trained well with training R2 above 0.85 and training RMSE below 75 MPa. RF exhibited the highest test accuracy with test R2 of 0.84 and test RMSE of 61 MPa, followed by XGBoost having test R2 of 0.74 and test RMSE of 77 MPa.

## 5.1. Experimental validation: Alloys printing, testing, and comparison with ML predictions

To evaluate the predictive and generalization capabilities of the best-performing ML model, the two alloys, $Co_{33.3}Cr_{33.3}Ni_{33.3}$ and $Co_{25}Cr_{45}Ni_{30}$, were printed using the LMD technique, and their YS were measured and compared. Notably, $Co_{33.3}Cr_{33.3}Ni_{33.3}$ was within the training dataset's composition range, while $Co_{25}Cr_{45}Ni_{30}$ was chosen to assess the model's extrapolation capabilities, as it was outside the training dataset's composition range. The printing parameters for both alloys were within the training dataset range.

XRD patterns acquired perpendicular to the build directions of printed alloys, shown in Figure 3a, reveal them to have a single-phase face-centered cubic (FCC) crystal structure, with peaks corresponding to the {111}, {200}, {220}, {311}, and {222} planes of the lattice. The {111} peaks showed higher intensities, indicating a slight <111> texture perpendicular to the build direction. Optical micrographs in Figure 3b and Figure 3c reveal melt-pool tracks, grain boundaries, and extensive dislocation cell structures inside grains in both alloys, with an average grain size varying between approximately 50 to 300 μm.

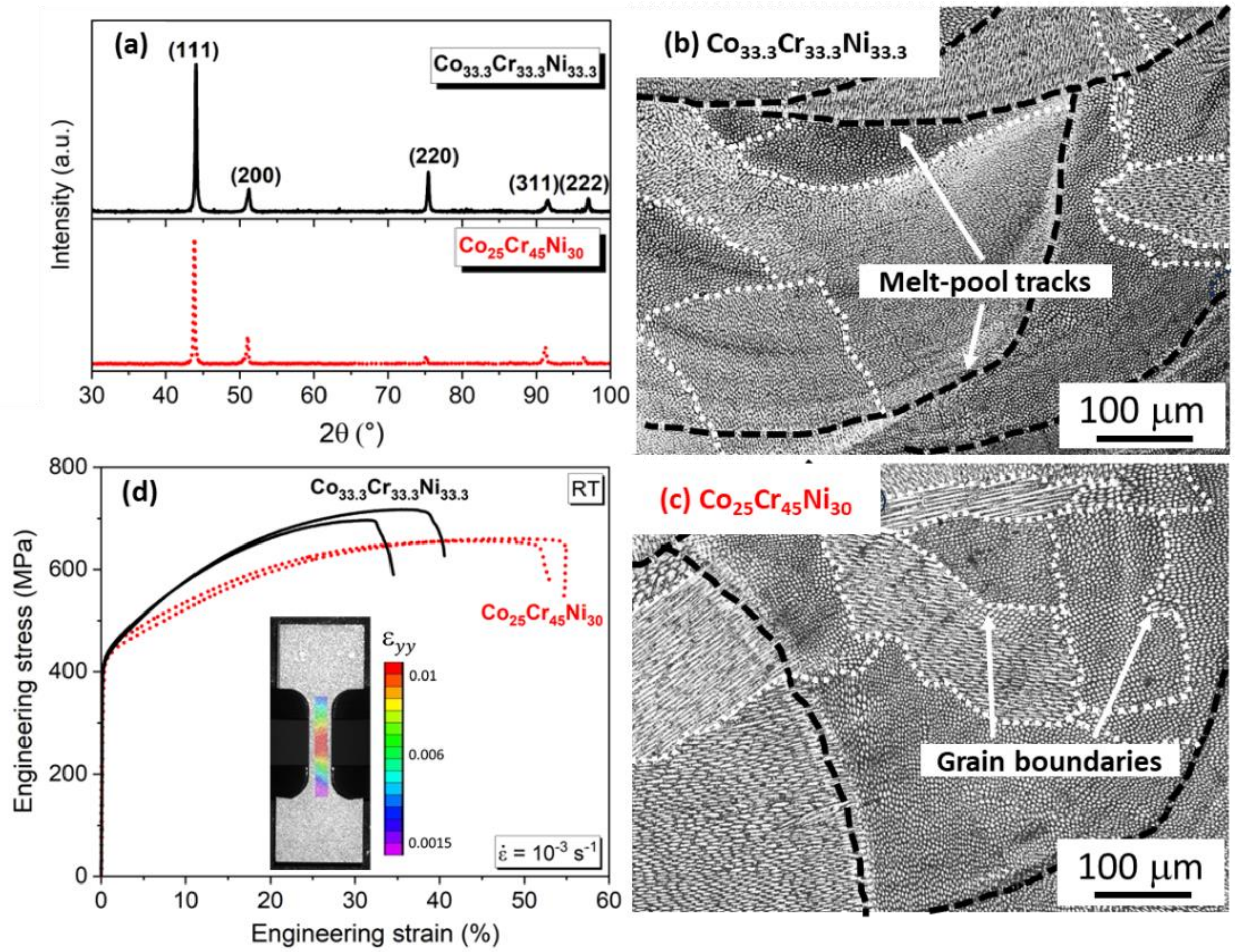

Figure 3: Experiment results from two printed and tested FCC MPEAs ($Co_{33.3}Cr_{33.3}Ni_{33.3}$and $Co_{25}Cr_{45}Ni_{30}$). (a) XRD patterns showing single-phase FCC peaks. Optical micrographs from (b) $Co_{33.3}Cr_{33.3}Ni_{33.3}$ and (c) $Co_{25}Cr_{45}Ni_{30}$ MPEAs showcase marked melt-pool tracks, grain boundaries, and extensive dislocation cell structures inside grains. (d) Engineering stress-strain curves obtained after tensile testing the two printed alloys at 293 K and $10^{-3}$ $s^{-1}$ strain rate. The tensile samples were speckle patterned randomly for DIC strain measurement, as shown in the inset. The overlayed strain map obtained using the Vic2D software illustrates the strain distribution within the gauge section. Both alloys show similar yield strengths but different work hardening rates and amounts, affecting ultimate tensile strength and total-elongation-to-failure.

The engineering stress-strain curves for the two printed alloys are shown in Figure 3d. Each test was repeated twice and showed reasonable repeatability. The curves indicate similar yield strength (stress at 0.2% strain offset) of approximately 416 MPa for both printed alloys ($Co_{33.3}Cr_{33.3}Ni_{33.3}$ and $Co_{25}Cr_{45}Ni_{30}$). However, different work hardening rates and magnitudes result in different ultimate tensile strengths and total-elongation-to-failures. The reasons for the differences and the similarities in the two alloys' responses are part of another study and will be reported separately. The comparison of the experimentally measured and RF-predicted YS for both alloys is shown in Table 3. Evidently, the estimated prediction errors of 59.7 MPa for $Co_{33.3}Cr_{33.3}Ni_{33.3}$ and 70.3 MPa for $Co_{25}Cr_{45}Ni_{30}$ are comparable to the RF model's RMSE on the testing dataset (61.0 MPa). This demonstrates the robustness and extrapolation capabilities of the developed ML model in predicting YS of compositions similar to the training dataset.

Table 3: Comparison of the experimentally measured and RF model predicted YS of two printed and tested FCC MPEAs ($Co_{33.3}Cr_{33.3}Ni_{33.3}$ and $Co_{25}Cr_{45}Ni_{30}$).

| Printed compositions | Experimental YS (MPa) | RF model predicted YS (MPa) | Prediction error (MPa) |
|---|---|---|---|
| **$Co_{33.3}Cr_{33.3}Ni_{33.3}$** | 416 ± 6 | 475.7 | 59.7 |
| **$Co_{25}Cr_{45}Ni_{30}$** | 416.5 ± 7 | 486.8 | 70.3 |

## 5.2. Comparison with microstructural information driven empirical predictions

As stated earlier, microstructural input features were bypassed while building ML models. However, to compare ML predictions against those obtained using empirical methods, data was gathered from the studies that reported consistent and complete microstructural information (see supplementary section 10). This microstructural data was then fed into standard empirical equations to estimate contributions from possible strengthening sources. Key fundamental strengthening mechanisms in single-phase MPEAs include lattice or solid solution strengthening, grain boundary strengthening, and dislocation forest strengthening. Other relevant strengthening contributions from nano-oxide particles[27,30,70,80] and elemental segregations at cell boundaries[92] were incorporated as per applicability. Since deformation twins typically form after yielding in some MPEAs during tensile testing[104], they were not considered as strengthening sources.

The following classical Hall–Petch equation[105] was employed for lattice and grain boundary strengthening.

$$\sigma_{HP} = \sigma_0 + \sigma_g = \sigma_0 + k \times d^{-\frac{1}{2}} \qquad \text{Equation 5}$$

where $\sigma_0$ is the lattice friction/solid solution contribution, $k$ is the Hall-Petch coefficient (MPa·μm$^{1/2}$), and $d$ is the mean grain size obtained from EBSD data. Some studies have extended the boundary-strengthening effect to include AM-induced sub-structures, using their size in the Hall-Petch equation[92,106]. This consideration is due to solute segregations at cell walls, which lead to dislocation pinning and additional strengthening[107,108].

The AM samples typically exhibit high dislocation densities to accommodate residual stress-induced strains from the high thermal gradients due to rapid cooling. These dislocations contribute to the dislocation forest strengthening. The Williamson-Hall method[109,110] is commonly employed to calculate dislocation density via XRD analysis[77]. Since dislocations are predominantly stored in the cellular walls of sub-grain structures, some studies[27] have utilized cell size for dislocation density estimation. The contribution of dislocations to strengthening was quantified by following Taylor's hardening relation[111].

$$\sigma_d = M \cdot \alpha \cdot G \cdot b \cdot \rho^{1/2} \qquad \text{Equation 6}$$

where $M$ is the Taylor factor, $\alpha$ is an obstacle constant, $G$ is shear modulus, $b$ is the burgers vector, and $\rho$ is the dislocation density[92].

Sub-micron-sized nano-oxide particle formation has been observed in some studies[27,30,70,80]. These oxides are manganese-rich and are more likely to form for the LPBF route. The formation of chromium-rich carbides has also been reported in cases where carbon was present in the alloy. The contribution from these particles was calculated by utilizing the inter-particle spacing (L) and mean particle diameter ($d_0$) according to the following Orowan's strengthening relation[112].

$$\sigma_p = \frac{0.4\,M}{\pi\sqrt{1-v}}\,\frac{Gb}{L}\ln\left(\sqrt{\frac{2}{3}}\,d_0\,/\,b\right) \qquad \text{Equation 7}$$

where $v$ denotes the Poisson's ratio. $L$ was calculated via the following equation:

$$L = \sqrt{\frac{2}{3}}\,d_0\,\sqrt{\frac{\pi}{4f} - 1} \qquad \text{Equation 8}$$

where $f$ is the volume fraction of particles. The microstructural details, materials-specific constants (e.g., lattice friction, Hall-Petch constant, obstacle constant, shear modulus, Burgers vector, Taylor factor), and estimated strengthening contributions are provided in the supplementary materials section 10. From Section 10 and Table S10 of the supplementary materials, a few general conclusions can be drawn. Table S10 highlights that the lattice friction/solid solution strengthening contributes significantly to the overall strength of these alloys, followed by dislocation forest strengthening and Hall-Petch strengthening. Notably, the Hall-Petch contribution is higher in LPBF-printed $Co_{33.3}Cr_{33.3}Ni_{33.3}$ alloys due to their lower grain size (Table S9). Precipitation hardening plays a significant role when oxide particles are added intentionally.

Various summation strategies of the above-estimated strengthening contributions are typically applied to estimate YS. Several reports have observed that linear summation overestimates the actual YS[113]. Consequently, many alternative superposition laws have been proposed, covering the entire range from root mean squared (RMS) to linear superposition. Therefore, the YS can fall between two bounds, with RMS as the lower bound and linear summation as the upper bound, as shown in Equation 9.

$$\sqrt{\sum_i \sigma_i^2} \leq \sigma_{0.2} \leq \sum_i \sigma_i \qquad \text{Equation 9}$$

where $\sigma_i$ is the strengthening contribution of the $i_{\text{th}}$ mechanism. Pythagorean superposition is particularly beneficial when considering phenomena operating at similar length scales. For instance, in AM printed parts, the mean spacing of dislocations and precipitates/segregations is expected to be comparable, both of which will act as short-range obstacles[114–116]. Meanwhile, grain size is much larger, so grain boundaries are long-range obstacles[114–116]. Since both dislocations and oxide particles will impede the motion of dislocations, their strengthening contributions ought to be merged and cannot be simply additive. This is generally carried out using Equation 10 and labelled as RMS-1 summation [114–116].

$$\sigma_y = \sigma_0 + \sigma_g + \sqrt{\sigma_d^2 + \sigma_p^2} \qquad \text{Equation 10}$$

The estimated YS for various MPEA compositions using linear, RMS-1, and RMS summations are provided in the supplementary materials section 10, while the comparison with the experimental YS is shown in Figure 4. The error bands representing the RF model test RMSE (61 MPa) and twice that value (122 MPa) are also compared. Evidently, the linear summation provided the best prediction accuracy with an $R^2$ of 0.81 and RMSE of 71 MPa, with most of the data points falling within the ± 122 MPa error band. The accuracy for RMS-1 summation was slightly lower with an $R^2$ of 0.73 and RMSE of 83 MPa, even though the same number of data points fall within the ± 122 MPa error band. Lastly, the RMS summation significantly underestimated the YS and showed poor accuracy, with most data points falling outside the ± 122 MPa error band.

The empirical predictions obtained using the linear summation, identified as the best empirical summation method, are compared with those from the best-performing ML-based RF model in Figure 5. Notably, the RF model predicted the experimental YS reasonably well, with higher $R^2$ and lower RMSE than the empirical linear summation model. Additionally, most data points from ML predictions fall within the ± 61 MPa error band. In comparison, several predicted via the empirical linear summation model fell outside the ± 61 MPa error band.

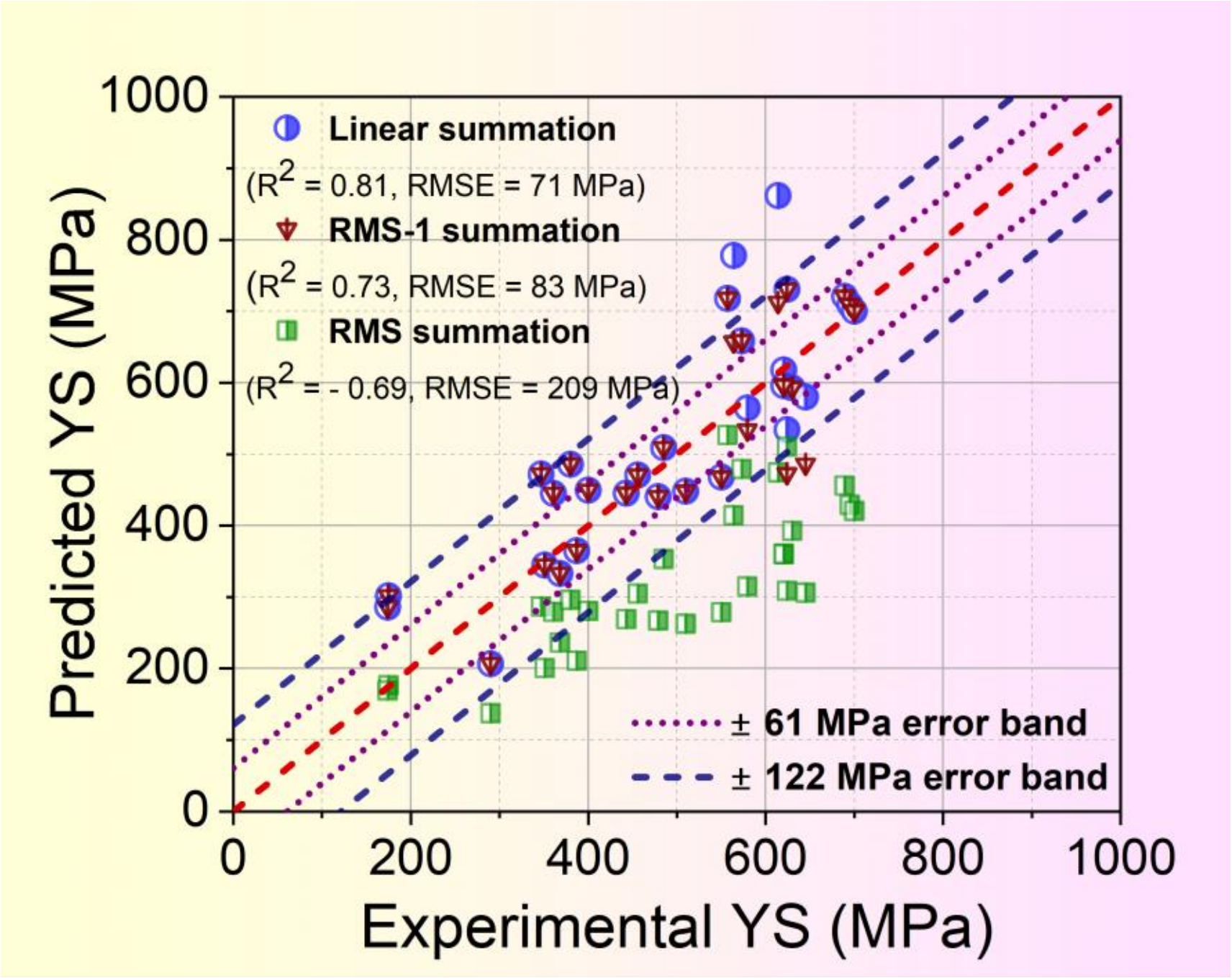

Figure 4: Comparison of the experimental YS with that predicted empirically using linear, RMS-1, and RMS summations. The error bands representing the RF model test RMSE (61 MPa) and twice that value (122 MPa) are also compared. Linear summation provided the best prediction accuracy with an $R^2$ of 0.81 and RMSE of 71 MPa. The accuracy for RMS-1 was slightly lower, with an $R^2$ of 0.73 and RMSE of 83 MPa. RMS summation significantly underestimated the strength and showed poor accuracy.

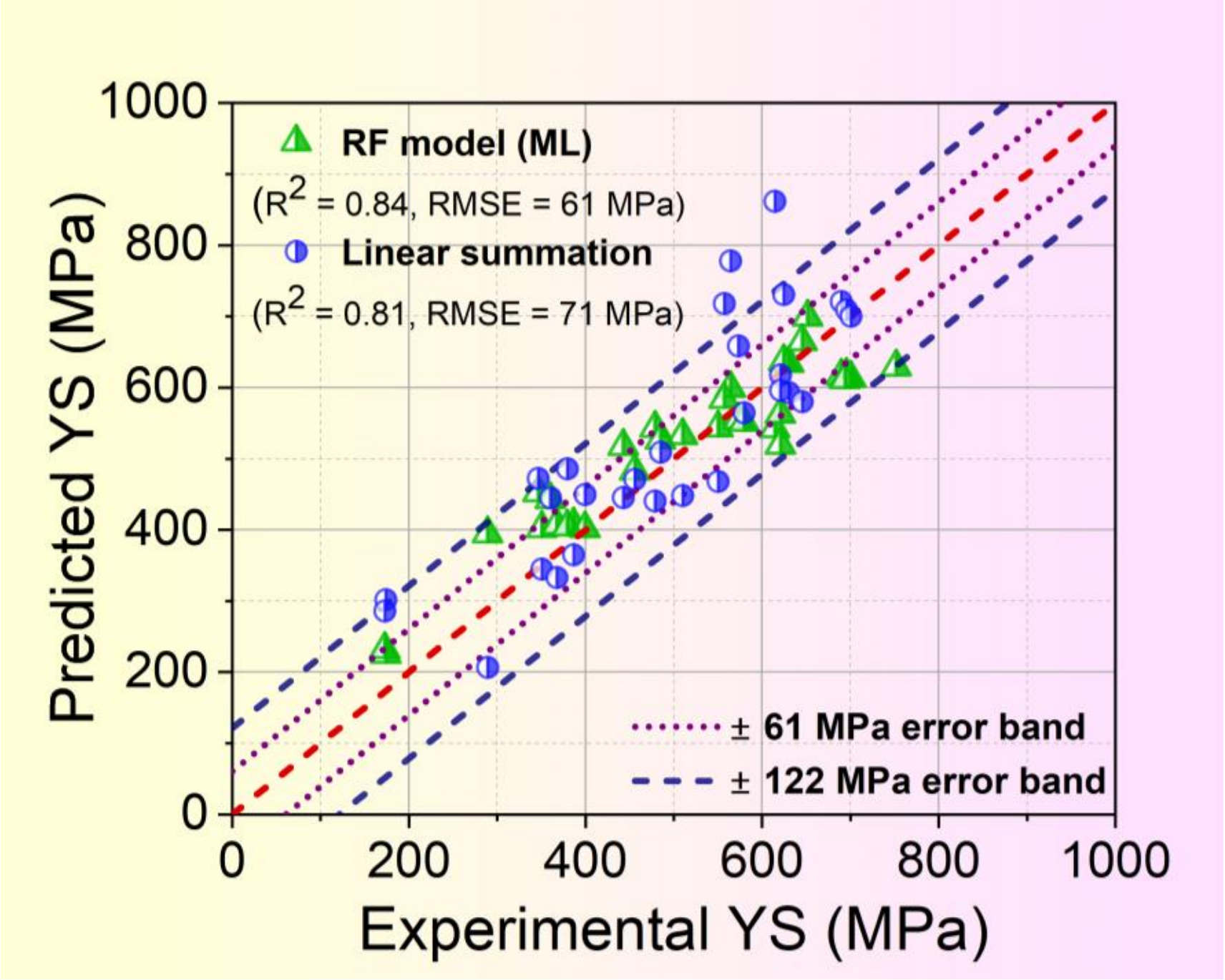

Figure 5: Comparison of the experimental YS with that predicted using empirical linear summation and best-performing ML-based RF model. The error bands representing the RF model test RMSE (61 MPa) and twice that value (122 MPa) are also compared. Notably, the ML-based RF model predicted the experimental YS reasonably well, with higher R2 and lower RMSE than the empirical linear summation predictions. Additionally, most data points from ML predictions lie within the ± 61 MPA error band, while several of those predicted via the empirical linear summation model fell outside the ± 61 MPa error band.

## 6. Analysis and discussion

### 6.1. Drawbacks of the empirical approach

As noted in the introduction, the empirical approach described above has several drawbacks. Briefly, empirical relations originally developed for conventionally manufactured alloys with relatively homogeneous microstructures may be overly simplistic for predicting the YS of AM materials. They generally fail to account for the hierarchical features, including irregularly shaped and textured grains, the high density of sub-grain boundaries, and elemental segregation at these boundaries. Grain size quantification is also challenging and lacks standardization for inhomogeneous and anisotropic printed microstructures, resulting in ambiguity in boundary-strengthening contributions. Moreover, as seen above, empirical estimations rely on precise materials-specific constants (e.g., lattice friction, Hall-Petch constant, obstacle constant, shear modulus, Burgers vector, Taylor factor), which are currently unavailable for novel compositions or have not been thoroughly measured. As a result, approximations from similar compositions are often used, potentially impacting the accuracy of empirical predictions.

Additionally, material constants, such as lattice friction and Hall-Petch constants, vary significantly in the literature, leading to ambiguity in empirical calculations. For example, Figure *6* shows the yield strength as a function of grain size from Yoshida et al.[5] and Schneider et al.[6], which estimated material constants for conventionally manufactured $Co_{33.3}Cr_{33.3}Ni_{33.3}$ alloy (see also Table *4*). The evident differences in their estimated lattice friction and Hall-Petch constants have not been critically examined before and could be due to several factors. For instance, Yoshida et al.[5] tensile specimens had cross-sections too small for large grain-sized variants (> 40 µm). In general, it is recommended to have a minimum of 10-12 grains[117] in the specimen cross-section for accurate material properties estimation. The impurity levels of the alloys used by Yoshida et al.[5] and Schneider et al.[6] may differ, potentially contributing to the variations in lattice friction and Hall-Petch constants, as highlighted by Schneider et al.[6]. Moreover, some studies [118] exclude twin boundaries from strengthening contributions, which can result in overestimated Hall-Petch constant values. For instance, Schneider et al.[6] found a 215 MPa $\mu m^{1/2}$ higher Hall-Petch constant when twin boundaries were excluded (see Table 4). This occurs because the twin boundaries in $Co_{33.3}Cr_{33.3}Ni_{33.3}$ alloy have been shown to possess strength similar to that of grain boundaries[6]. Consequently, if these twin boundaries are excluded, the Hall-Petch constant in Equation 6 increases to adjust for the larger grain size used in the calculations[5][6].

Furthermore, most studies assume that grains are free of dislocations after annealing. However, residual dislocations from prior deformation may persist, depending on the deformation extent and suboptimal annealing. These remnant dislocations can contribute significantly to strengthening, especially in smaller grain sizes, and should be accounted for, as they may influence both the lattice friction and Hall-Petch constants.

Yin et al.[118] also analyzed data on conventionally manufactured $Co_{33.3}Cr_{33.3}Ni_{33.3}$ and found it consistent with the Hall-Petch model, reporting lattice friction and Hall-Petch constants of 218 MPa and 537 MPa $\mu m^{1/2}$, respectively. However, their analysis does not consider the data points from the study by Schneider et al.[6], which limits its acceptability. Additionally, both Schneider et al.[6] and Yin et al. [118] reported good agreement between their calculated lattice friction strengths and theoretical solid solution strengthening models, even though their calculated values were different. These discrepancies indicate the need for further investigation.

Despite these variations, the material constants reported by Yoshida et al.[5] have been majorly employed for microstructural information-driven yield strength predictions for AM $Co_{33.3}Cr_{33.3}Ni_{33.3}$ alloy (Figure 5 and Figure 6). However, when alternative values are used, significant declines in prediction accuracy (significantly lower $R^2$ and higher RMSE) are observed (see Table 4). Therefore, accurate material constant values are crucial for reliable empirical predictions.

Table 4: Lattice friction and Hall-Petch constant values reported for conventionally manufactured $Co_{33.3}Cr_{33.3}Ni_{33.3}$ alloy and the resulting overall accuracies for linear summation empirical predictions.

| **$Co_{33.3}Cr_{33.3}Ni_{33.3}$ MPEA** | **Lattice friction (MPa)** | **Hall-Petch constant (MPa $\mu m^{1/2}$)** | **$R^2$** | **RMSE** |
|---|---|---|---|---|
| **Yoshida et al.[5] (including twin boundaries)** | 218 | 265 | 0.81 | 71 |
| **Schneider et al.[6] (including twin boundaries)** | 135 | 600 | 0.25 | 140 |
| **Schneider et al.[6] (excluding twin boundaries)** | 150 | 815 | -1.2 | 237 |
| **Yin et al. [118] (excluding twin boundaries)** | 218 | 537 | 0.10 | 152 |

The ambiguity in grain size quantification could also contribute to reduced accuracies in empirical model predictions. While most studies utilized the line intercept method to measure the arithmetic average grain size, few have considered EBSD software analysis[52,54,84]. To account for the heterogeneous grain sizes, some have applied alternate approaches like volume-weighted average[48], rule of mixture[41], and modified Hall-Petch relation[86] for crescent-shaped grains. Additionally, some studies have expanded the boundary-strengthening effect to account for AM-induced sub-grain structures by incorporating their size into the Hall-Petch equation[92,106]. This approach accounts for solute segregation at cell walls, causing dislocation pinning and additional strengthening[107,108]. However, it likely results in a significant overestimation, as low-angle grain boundaries are treated as high-angle grain boundaries in the applied Hall-Petch model

Moreover, empirical yield strength predictions rely only on the as-printed microstructure, overlooking the initial printing parameters. Given these limitations, a data-driven approach presents a promising alternative.

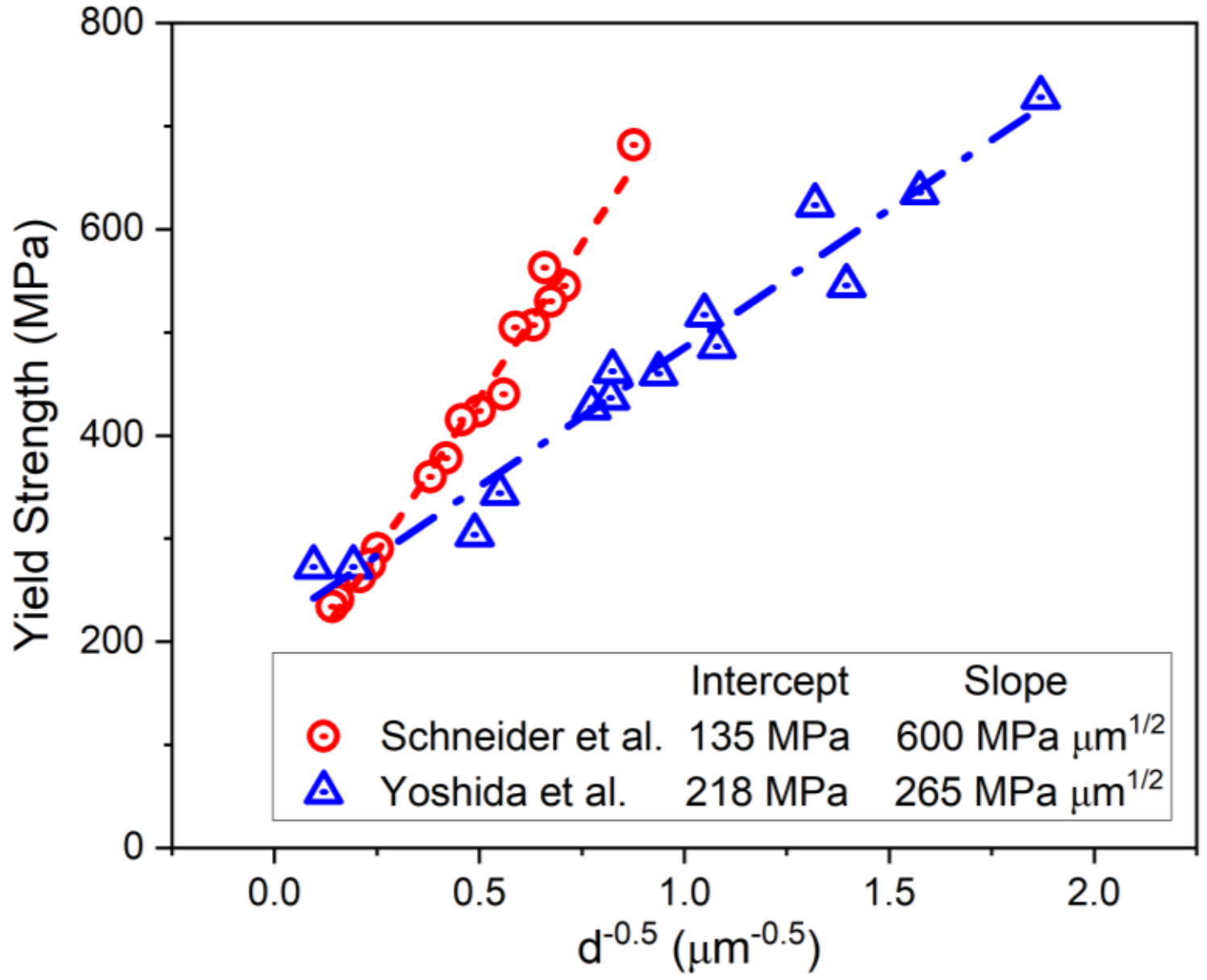

Figure 6. Yield strength of conventionally manufactured $Co_{33.3}Cr_{33.3}Ni_{33.3}$ as a function of grain size reported by Yoshida et al.[5] and Schneider et al.[6]. The grain sizes, estimated by including twin boundaries, vary from 0.199 nm to 111 μm for Yoshida et al.[5] and 3.2 to 174 μm for Schneider et al.[6]. The intercept and slope give the value of lattice friction and Hall-Petch constant, respectively. Schneider et al.[6] reported 83 MPa lower lattice friction and 335 MPa μm$^{1/2}$ higher Hall-Petch constant compared to Yoshida et al.[5].

## 6.2. Input feature importance

Post-hoc analysis was conducted using SHAP for the best-performing RF model. This analysis determined the importance of the hierarchy of input features and visualized the effect of individual input feature values on the predicted YS. Since the two manufacturing routes (LPBF and LMD) have different setups, their importance hierarchies are expected to differ. Additionally, the ranges of manufacturing route-related parameter values, such as laser power and scan velocity, vary significantly, as shown in the supplementary material section 2. Consequently, the importance hierarchy was analyzed separately for the LPBF and LMD data, with SHAP results in Figure 7.

For both routes, printing-related features such as scan pattern rotation, energy density, laser power, and scan velocity were the most important, followed by testing-related parameters such as cross-section area, test temperature, and test direction. Composition-related variables such as atomic weight mismatch, mean atomic weight, Peierls Nabarro factor, and modulus mismatch in the strengthening model, were next in importance. Each data point of an input feature is represented by a colored dot, indicating the magnitude of the feature value, while its x-axis position indicates its corresponding SHAP value. The SHAP value measures the relative contribution of the feature value of a data point to the predicted outcome. A more positive SHAP value indicates a more positive effect of the feature value on the predicted value and vice versa.

It is evident that the scan pattern rotation significantly influences the YS. Higher feature values (colored in red) for the scan pattern rotation exhibit higher SHAP values, indicating a strong positive influence on YS. These higher encoded values correspond to scanning with rotation after every layer, while the lower values correspond to scanning without rotation and bidirectional scanning. Scanning without rotation results in coarser and more columnar grain structures than with rotation after every layer[26,28,119]. This is due to the nearly constant direction of the temperature gradient throughout the printing process, facilitating the growth of large columnar grains[120], which likely explains the observed SHAP result. Similarly, lower scan velocity is associated with a lower predicted YS. Conversely, input features such as laser power, LMD linear heat input, and SLM VED exhibit an inverse relationship with YS. This is expected because a lower scan rate and higher laser power result in higher heat input to the material[1]. Consequently, the previously solidified regions surrounding the laser spot and the lower layers experience prolonged exposure to elevated temperatures[3]. This results in recovery in the as-printed microstructure, including grain coarsening, annihilation of dislocations in the cellular structures, and reduction in residual stresses, ultimately lowering the YS[120].

An increase in testing temperature has been observed to cause a decrease in YS. This could be due to the activation of several mechanisms[69], including (1) the annihilation of forest dislocations at high temperatures, reducing the obstruction to dislocation movement, (2) the facilitation of overcoming short-range obstacles due to higher thermal energy, (3) the reduction of Peierls lattice friction stress due to lowering of the shear modulus, and (4) the activation of the dislocation climb mechanism, enabling the migration of edge dislocations. Figure 7 demonstrates the significance of the specimen cross-section area as an important feature. However, further investigations are required, particularly in the context of AM-based microstructure length scale-driven mechanical testing, to confirm the importance of this feature[121]. Strain rate, in general, influences the mechanical properties of materials; however, its low importance in the hierarchy list may be attributed to most of the tests reported in the literature being performed at quasi-static conditions. The lack of variation in strain rate values makes it difficult to capture its effect.

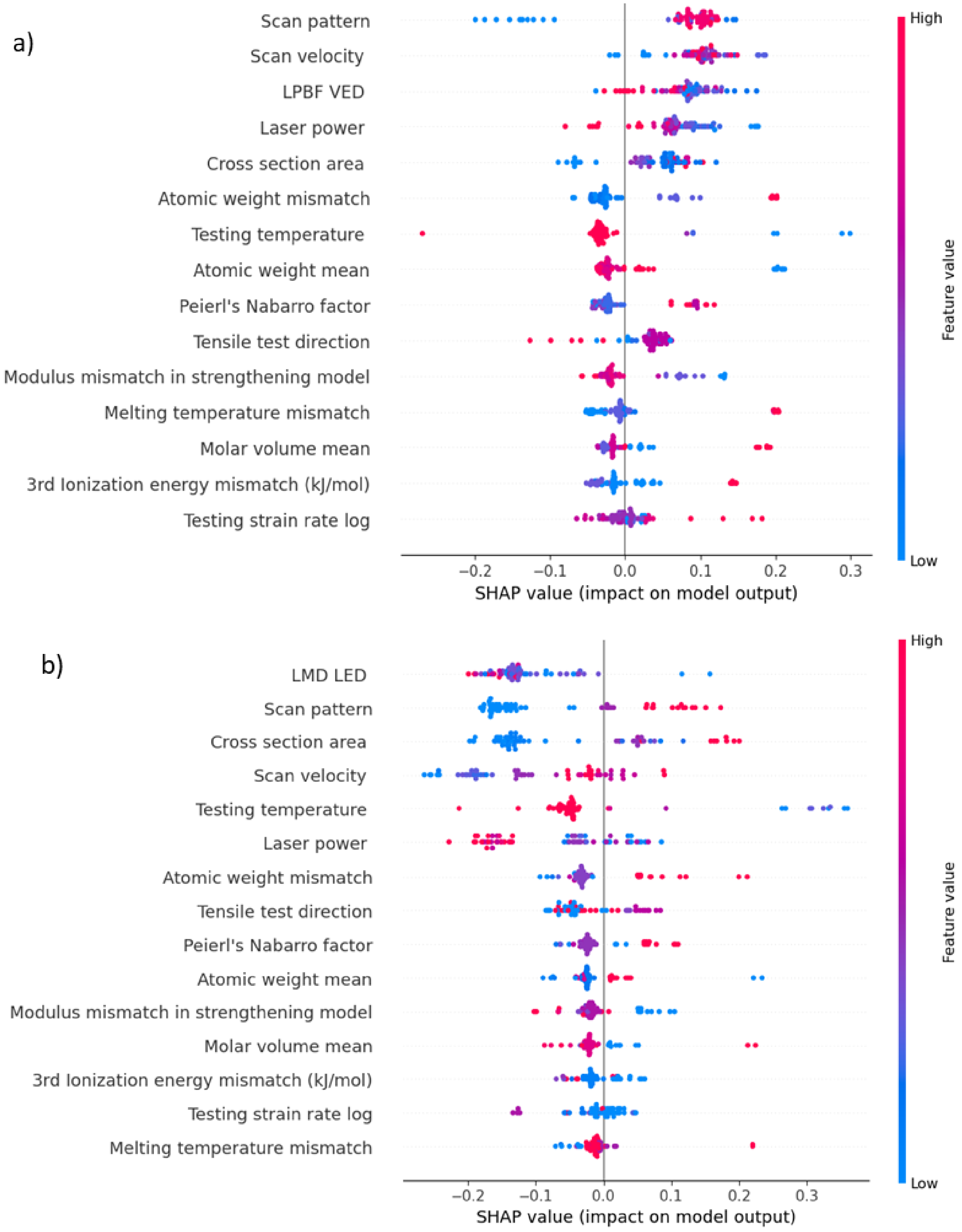

Figure 7: SHAP analysis plots showcasing input features importance hierarchy for the RF model considering the alloys fabricated via a) LPBF and b) LMD routes. Data points are depicted as colored dots, with the color indicating the magnitude of the feature values (red being high and blue being low). The x-axis's sign (positive or negative) and magnitude indicate the relative effect of the feature values on the predicted YS.

Among the compositional input features, the mismatch of atomic weight has the highest importance and is linked to higher YS. Higher chromium content and the addition of carbon are associated with larger mismatch values due to their lower atomic weights than other elements. This aligns with the ability of chromium to induce greater mean atomic displacement in the lattice of MPEAs and the interstitial strengthening effect of carbon[30,122]. Both elements enhance material strength by introducing geometrical lattice distortion and generating stress fields due to local changes in shear modulus. These stress fields interact with dislocation stress fields[123] and impede their motion. The Peierls-Nabarro factor of the compositions also positively affects YS. A higher Peierls-Nabarro factor is associated with higher shear modulus values, indicating stronger stress fields and higher lattice friction, contributing to strengthening.

The input feature trends are similar for both LPBF and LMD processes, though the magnitude varies. This is due to the comparable melting and solidification processes in both methods, which result in similar effects of parameter variations on microstructure and properties.

## 6.3. Feature trends

Feature trends were produced to analyze the effects of individual printing parameters on YS prediction. A separate synthetic dataset was prepared for this analysis, with variations only in the investigated parameter. To ensure reliable predictions, the data points in the synthetic dataset were kept within the limits of the dataset used for model training. The most common composition, $Co_{20}Cr_{20}Fe_{20}Ni_{20}Mn_{20}$, and a tensile testing strain rate of $10^{-3}$ $s^{-1}$ at room temperature were considered. To reduce bias from the specific feature weights in a model and to improve generalizability, it is advisable to consider multiple models[124]. Therefore, the two best models from previous development were used to predict the YS for the synthetic dataset, and their mean was also considered.

Despite some scatter, an overall increase in predicted strength is observed for increasing scan velocities (Figure 8a). This aligns with the SHAP plot results and can be attributed to the decrease in grain size due to faster solidification rates. Conversely, an overall reduction in strength is predicted for increasing laser power (Figure 8b), which is consistent with the SHAP results and has also been experimentally observed. For example, Wang et al.[64] and Tong et al.[125] reported a reduction in the YS with increasing laser powers for $Co_{20}Cr20Fe_{20}Mn_{20}Ni_{20}$ fabricated via the LPBF route. Xue et al.[94] also observed the same effect for $Co_{33.3}Cr_{33.3}Ni_{33.3}$ fabricated via the LMD route. They reported increased cell structure size with increasing laser power, and the associated decrease in the dislocation density might have contributed to the YS reduction. The effect of the scan pattern rotation on YS prediction is shown in Figure 8c. The predicted YS is lowest with no rotation, increases with a 180° rotation between layers, and is highest with 67° and 90° rotations. This observation is also reflected in the SHAP plot, which can be explained by the disruption of heat gradients between layers due to scan pattern rotation, which inhibits columnar grain growth and promotes refined equiaxed grains.

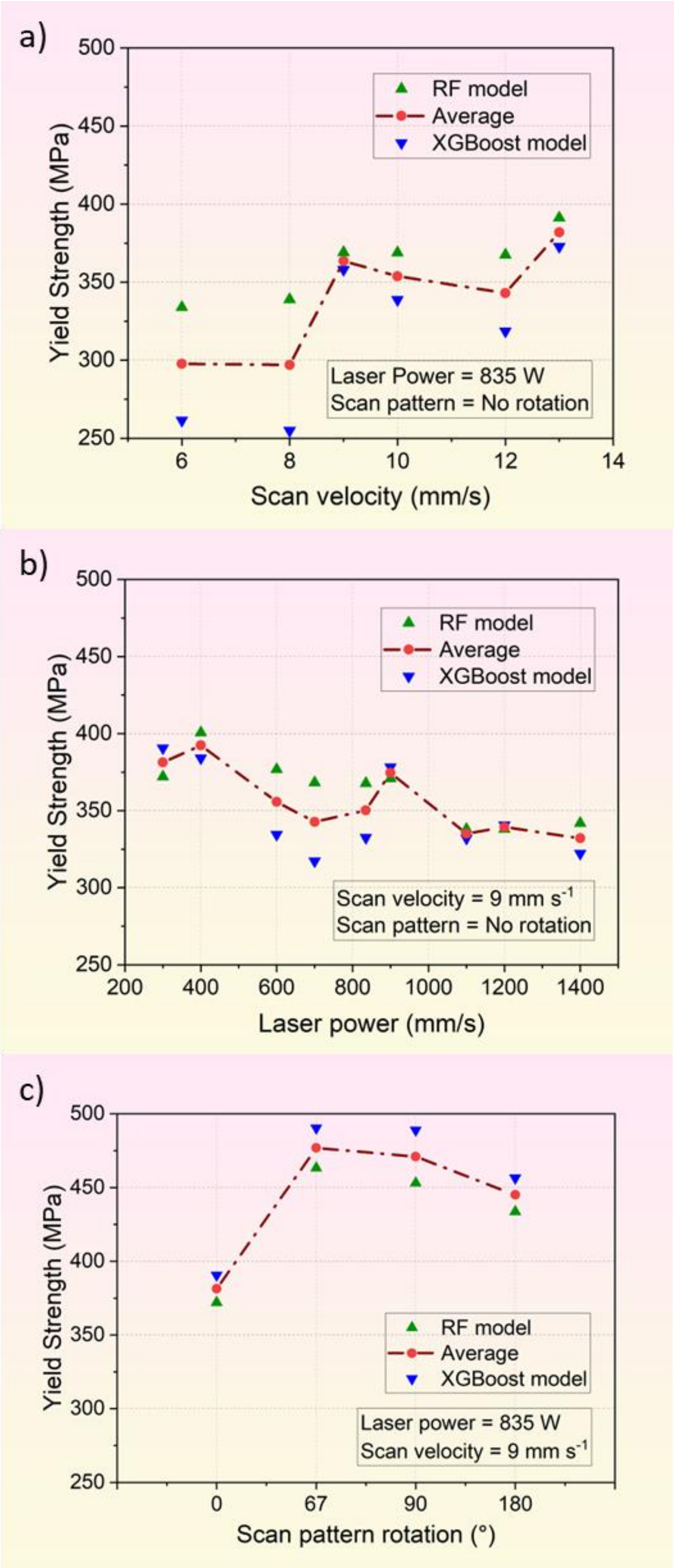

Figure 8: Trends obtained by plotting RF and XGBoost models predicted YS for the synthetic dataset of $Co_{20}Cr_{20}Fe_{20}Mn_{20}Ni_{20}$ alloy printed via LMD technique as a function of a) scan velocity, b) laser power, and c) scan pattern rotation. While scan velocity positively affects the predicted YS, laser power exhibits a relatively inverse effect. The predicted YS is lowest for no rotation, increases for a 180° rotation between layers, and is highest for 67° and 90° rotations.

## 7. Summary and conclusions

This study proposes an experimentally validated and empirically compared ML approach for predicting the YS of additively manufactured MPEAs from the Co-Cr-Fe-Mn-Ni system. The trained and tested RF, XGBoost, and GBoost algorithms demonstrated reasonable accuracy, with the RF model performing the best, achieving a test $R^2$ of 0.84 and test RMSE of 61.0 MPa. The RF model successfully predicted the YS of two printed and tested alloys, one from the dataset ($Co_{33.3}Cr_{33.3}Ni_{33.3}$) and one novel ($Co_{25}Cr_{45}Ni_{30}$), with reasonable accuracy showcasing the robustness and extrapolation capabilities of our approach. The RF model also performed on par with the empirical predictions obtained using microstructural detail-driven strengthening contributions while having the advantage of establishing the effect of initial printing parameters on the YS. Post-hoc analysis of the developed model revealed the significance of various input features on the ML-predicted YS, corroborated by constructing feature trends. The key takeaways are:

a) Rotation in the scan pattern is associated with higher YS values, attributed to the disruptive effect of rotation on the temperature gradient, inhibiting columnar grain growth.
b) Scan velocity exhibits a positive correlation with YS. Conversely, laser power and energy input show an inverse relationship with YS, as higher heat input facilitates recovery, alleviates residual stresses, and leads to grain coarsening.
c) Among the composition descriptors, the most important input features were mismatch in atomic weight, mean atomic weight and Peierls-Nabarro factor. This is possibly linked to the presence of chromium and carbon in the MPEAs composition, which introduces stress fields for dislocations to interact with and obstruct their motion, leading to strengthening.

Overall, this study demonstrates that quick and accurate property prediction is achievable through a data-driven methodology, eliminating the need for time-consuming and resource-intensive processes to gather microstructural information for iterative experiments. Additionally, this approach provides significant value by elucidating the effects of parameters and guiding the tuning of mechanical properties of additively manufactured MPEAs. By expanding the dataset, collecting more structured data, incorporating additional parameters, and integrating more advanced ML techniques, it is anticipated that predicting other mechanical properties for a broader range of compositions with multiple phases will become possible.

### Acknowledgments

AC acknowledges financial supports from the Infosys Foundation and the Indian Institute of Science, Bengaluru. The first author receives financial support through the PMRF scheme. The authors would like to acknowledge and express their gratitude to Prof. Punit Rathore from Robert Bosch Centre for Cyberphysical Systems, IISc, Bengaluru, for his valuable insights and discussions on ML algorithms.

# References

1 M.A. Buhairi, F.M. Foudzi, F.I. Jamhari, A.B. Sulong, N.A.M. Radzuan, N. Muhamad, I.F. Mohamed, A.H. Azman, W.S.W. Harun, and M.S.H. Al-Furjan: *Progress in Additive Manufacturing*, 2022.

2 T. Ikeda, M. Yonehara, T.T. Ikeshoji, T. Nobuki, M. Hatate, K. Kuwabara, Y. Otsubo, and H. Kyogoku: *Crystals (Basel)*, DOI:10.3390/cryst11050549.

3 M. Dada, P. Popoola, N. Mathe, S. Pityana, and S. Adeosun: in *Materials Today: Proceedings*, vol. 38, Elsevier Ltd, 2021, pp. 756–61.

4 Y. Liu, J. Ren, S. Guan, C. Li, Y. Zhang, S. Muskeri, Z. Liu, D. Yu, Y. Chen, K. An, Y. Cao, W. Liu, Y. Zhu, W. Chen, S. Mukherjee, T. Zhu, and W. Chen: *Acta Mater*, DOI:10.1016/j.actamat.2023.118884.

5 S. Yoshida, T. Bhattacharjee, Y. Bai, and N. Tsuji: *Scr Mater*, 2017, vol. 134, pp. 33–6.

6 M. Schneider, E.P. George, T.J. Manescau, T. Záležák, J. Hunfeld, A. Dlouhý, G. Eggeler, and G. Laplanche: *Int J Plast*, 2020, vol. 124, pp. 155–69.

7 N. Kouraytem, X. Li, W. Tan, B. Kappes, and A.D. Spear: *Journal of Physics: Materials*, 2021, vol. 4, p. 032002.

8 S. Liu, A.P. Stebner, B.B. Kappes, and X. Zhang: *Addit Manuf*, 2021, vol. 39, p. 101877.

9 P. Sreeramagiri and G. Balasubramanian: *Additive Manufacturing Letters*, 2022, vol. 3, p. 100045.

10 L.K. Chang, R.S. Chen, M.C. Tsai, R.M. Lee, C.C. Lin, J.C. Huang, T.W. Chang, and M.H. Horng: *International Journal of Advanced Manufacturing Technology*, 2024, vol. 132, pp. 83–98.

11 C. Herriott and A.D. Spear: *Comput Mater Sci*, DOI:10.1016/j.commatsci.2020.109599.

12 Z. Zhan and H. Li: *Int J Fatigue*, DOI:10.1016/j.ijfatigue.2020.105941.

13 I.Z. Era, M. Grandhi, and Z. Liu: *International Journal of Advanced Manufacturing Technology*, 2022, vol. 121, pp. 2445–59.

14 S.A. Rahman, A. Chandraker, O. Prakash, and A. Chauhan: *Eng Fract Mech*, 2024, vol. 306, p. 110214.

15 J. Xiong, S.Q. Shi, and T.Y. Zhang: *J Mater Sci Technol*, 2021, vol. 87, pp. 133–42.

16 L. Qiao, Z. Lai, Y. Liu, A. Bao, and J. Zhu: *J Alloys Compd*, DOI:10.1016/j.jallcom.2020.156959.

17 R. Jain, S.K. Dewangan, P. Umre, V. Kumar, and S. Samal: *Transactions of the Indian Institute of Metals*, 2021, vol. 74, pp. 2671–9.

18 S.K. Dewangan, S. Samal, and V. Kumar: *J Alloys Compd*, DOI:10.1016/j.jallcom.2020.153766.

19 K. Lee and P. V. Balachandran: *Materialia (Oxf)*, DOI:10.1016/j.mtla.2022.101628.

20 C. Wen, Y. Zhang, C. Wang, D. Xue, Y. Bai, S. Antonov, L. Dai, T. Lookman, and Y. Su: *Acta Mater*, 2019, vol. 170, pp. 109–17.

21 C. Yang, C. Ren, Y. Jia, G. Wang, M. Li, and W. Lu: *Acta Mater*, DOI:10.1016/j.actamat.2021.117431.

22 N.J. Sai, P. Rathore, and A. Chauhan: *Scr Mater*, DOI:10.1016/j.scriptamat.2022.115214.

23 P. Akbari, M. Zamani, and A. Mostafaei: .

24 Y.-K. Kim, M.-S. Baek, S. Yang, and K.-A. Lee: *Addit Manuf*, 2021, vol. 38, p. 101832.

25 R. Li, P. Niu, T. Yuan, P. Cao, C. Chen, and K. Zhou: *J Alloys Compd*, 2018, vol. 746, pp. 125–34.

26 B. Dovgyy, A. Piglione, P.A. Hooper, and M.S. Pham: *Mater Des*, DOI:10.1016/j.matdes.2020.108845.

27 P. Chen, C. Yang, S. Li, M.M. Attallah, and M. Yan: *Mater Des*, 2020, vol. 194, p. 108966.

28 M. Jin, A. Piglione, B. Dovgyy, E. Hosseini, P.A. Hooper, S.R. Holdsworth, and M.S. Pham: *Addit Manuf*, DOI:10.1016/j.addma.2020.101584.

29 Y.-K. Kim, J.-Y. Suh, and K.-A. Lee: *Materials Science and Engineering: A*, 2020, vol. 796, p. 140039.

30 J.M. Park, J. Choe, J.G. Kim, J.W. Bae, J. Moon, S. Yang, K.T. Kim, J.H. Yu, and H.S. Kim: *Mater Res Lett*, 2020, vol. 8, pp. 1–7.

31 H. Chen, T. Lu, Y. Wang, Y. Liu, T. Shi, K.G. Prashanth, and K. Kosiba: *Materials Science and Engineering: A*, 2022, vol. 833, p. 142512.

32 B. Li, L. Zhang, Y. Xu, Z. Liu, B. Qian, and F. Xuan: *Powder Technol*, 2020, vol. 360, pp. 509–21.

33 Y. Brif, M. Thomas, and I. Todd: *Scr Mater*, 2015, vol. 99, pp. 93–6.

34 Z. Sun, X. Tan, C. Wang, M. Descoins, D. Mangelinck, S.B. Tor, E.A. Jägle, S. Zaefferer, and D. Raabe: *Acta Mater*, 2021, vol. 204, p. 116505.

35 D. Lin, L. Xu, H. Jing, Y. Han, L. Zhao, and F. Minami: *Addit Manuf*, 2020, vol. 32, p. 101058.

36 Z.G. Zhu, Q.B. Nguyen, F.L. Ng, X.H. An, X.Z. Liao, P.K. Liaw, S.M.L. Nai, and J. Wei: *Scr Mater*, 2018, vol. 154, pp. 20–4.

37 D. Lin, L. Xu, H. Jing, Y. Han, L. Zhao, Y. Zhang, and H. Li: *Addit Manuf*, 2020, vol. 36, p. 101591.

38 D. Lin, L. Xu, Y. Han, Y. Zhang, H. Jing, L. Zhao, and F. Minami: *Intermetallics (Barking)*, 2020, vol. 127, p. 106963.

39 Y.O. Kuzminova, D.G. Firsov, S.A. Dagesyan, S.D. Konev, S.N. Sergeev, A.P. Zhilyaev, M. Kawasaki, I.S. Akhatov, and S.A. Evlashin: *J Alloys Compd*, 2021, vol. 863, p. 158609.

40 X.-H. Gu, T. Lu, T. Zhang, W. Guo, Y. Pan, and T. Dai: *Rare Metals*, 2022, vol. 41, pp. 2047–54.

41 J.M. Park, P. Asghari-Rad, A. Zargaran, J.W. Bae, J. Moon, H. Kwon, J. Choe, S. Yang, J.-H. Yu, and H.S. Kim: *Acta Mater*, 2021, vol. 221, p. 117426.

42 Z. Sun, X.P. Tan, M. Descoins, D. Mangelinck, S.B. Tor, and C.S. Lim: *Scr Mater*, 2019, vol. 168, pp. 129–33.

43 R. Zhou, Y. Liu, C. Zhou, S. Li, W. Wu, M. Song, B. Liu, X. Liang, and P.K. Liaw: *Intermetallics (Barking)*, 2018, vol. 94, pp. 165–71.

44 G.M. Karthik, Y. Kim, E.S. Kim, A. Zargaran, P. Sathiyamoorthi, J.M. Park, S.G. Jeong, G.H. Gu, A. Amanov, T. Ungar, and H.S. Kim: *Addit Manuf*, 2022, vol. 59, p. 103131.

45 H. Chen, K. Kosiba, T. Lu, N. Yao, Y. Liu, Y. Wang, K.G. Prashanth, and C. Suryanarayana: *J Mater Sci Technol*, 2023, vol. 136, pp. 245–59.

46 Z. Fu, B. Yang, K. Gan, D. Yan, Z. Li, G. Gou, H. Chen, and Z. Wang: *Corros Sci*, 2021, vol. 190, p. 109695.

47 Y.-T. Lin, X. An, Z. Zhu, M.L.S. Nai, C.-W. Tsai, and H.-W. Yen: *J Alloys Compd*, 2022, vol. 925, p. 166735.

48 C. Zhang, K. Feng, H. Kokawa, and Z. Li: *Materials Science and Engineering: A*, 2022, vol. 855, p. 143920.

49 R. LI, P. NIU, T. YUAN, and Z. LI: *Transactions of Nonferrous Metals Society of China*, 2021, vol. 31, pp. 1059–73.

50 Y. Peng, C. Jia, L. Song, Y. Bian, H. Tang, G. Cai, and G. Zhong: *Intermetallics (Barking)*, 2022, vol. 145, p. 107557.

51 R. Sokkalingam, Z. Chao, K. Sivaprasad, V. Muthupandi, J. Jayaraj, P. Ramasamy, J. Eckert, and K.G. Prashanth: *Adv Eng Mater*, DOI:10.1002/adem.202200341.

52 B. Han, C. Zhang, K. Feng, Z. Li, X. Zhang, Y. Shen, X. Wang, H. Kokawa, R. Li, Z. Wang, and P.K. Chu: *Materials Science and Engineering: A*, 2021, vol. 820, p. 141545.

53 H. Duan, B. Liu, A. Fu, J. He, T. Yang, C.T. Liu, and Y. Liu: *J Mater Sci Technol*, 2022, vol. 99, pp. 207–14.

54 J. Wang, J. Zou, H. Yang, L. Zhang, Z. Liu, X. Dong, and S. Ji: *J Mater Sci Technol*, 2022, vol. 127, pp. 61–70.

55 Z. Ma, Q. Zhai, K. Wang, G. Chen, X. Yin, Q. Zhang, L. Meng, S. Wang, and L. Wang: *Journal of Materials Research and Technology*, 2022, vol. 16, pp. 899–911.

56 Z. Qiu, C. Yao, K. Feng, Z. Li, and P.K. Chu: *International Journal of Lightweight Materials and Manufacture*, 2018, vol. 1, pp. 33–9.

57 S. Xiang, H. Luan, J. Wu, K.-F. Yao, J. Li, X. Liu, Y. Tian, W. Mao, H. Bai, G. Le, and Q. Li: *J Alloys Compd*, 2019, vol. 773, pp. 387–92.

58 Z. Tong, H. Liu, J. Jiao, W. Zhou, Y. Yang, and X. Ren: *Addit Manuf*, 2020, vol. 35, p. 101417.

59 X. Gao, Z. Yu, W. Hu, Y. Lu, Z. Zhu, Y. Ji, Y. Lu, Z. Qin, and X. Lu: *J Alloys Compd*, 2020, vol. 847, p. 156563.

60 M.A. Melia, J.D. Carroll, S.R. Whetten, S.N. Esmaeely, J. Locke, E. White, I. Anderson, M. Chandross, J.R. Michael, N. Argibay, E.J. Schindelholz, and A.B. Kustas: *Addit Manuf*, 2019, vol. 29, p. 100833.

61 Z. Tong, X. Ren, J. Jiao, W. Zhou, Y. Ren, Y. Ye, E.A. Larson, and J. Gu: *J Alloys Compd*, 2019, vol. 785, pp. 1144–59.

62 S. Guan, D. Wan, K. Solberg, F. Berto, T. Welo, T.M. Yue, and K.C. Chan: *Materials Science and Engineering: A*, 2019, vol. 761, p. 138056.

63 X. Zhang, R. Li, L. Huang, A. Amar, C. Wu, G. Le, X. Liu, D. Guan, G. Yang, and J. Li: *Vacuum*, 2021, vol. 187, p. 110111.

64 Y.L. Wang, L. Zhao, D. Wan, S. Guan, and K.C. Chan: *Materials Science and Engineering: A*, 2021, vol. 825, p. 141871.

65 Y. Bai, H. Jiang, K. Yan, M. Li, Y. Wei, K. Zhang, and B. Wei: *J Mater Sci Technol*, 2021, vol. 92, pp. 129–37.

66 X. Gao and Y. Lu: *Mater Lett*, 2019, vol. 236, pp. 77–80.

67 J. Li, H. Luan, L. Zhou, A. Amar, R. Li, L. Huang, X. Liu, G. Le, X. Wang, J. Wu, and C. Jiang: *Mater Des*, 2020, vol. 195, p. 108999.

68 Y. Chew, G.J. Bi, Z.G. Zhu, F.L. Ng, F. Weng, S.B. Liu, S.M.L. Nai, and B.Y. Lee: *Materials Science and Engineering: A*, 2019, vol. 744, pp. 137–44.

69 J. Li, S. Xiang, H. Luan, A. Amar, X. Liu, S. Lu, Y. Zeng, G. Le, X. Wang, F. Qu, C. Jiang, and G. Yang: *J Mater Sci Technol*, 2019, vol. 35, pp. 2430–4.

70 Y. Chew, Z.G. Zhu, F. Weng, S.B. Gao, F.L. Ng, B.Y. Lee, and G.J. Bi: *J Mater Sci Technol*, 2021, vol. 77, pp. 38–46.

71 L. Huang, Y. Sun, A. Amar, C. Wu, X. Liu, G. Le, X. Wang, J. Wu, K. Li, C. Jiang, and J. Li: *Vacuum*, 2021, vol. 183, p. 109875.

72 Y. Cai, M. Shan, Y. Cui, S.M. Manladan, X. Lv, L. Zhu, D. Sun, T. Wang, and J. Han: *J Alloys Compd*, 2021, vol. 887, p. 161323.

73 K. Zhou, J. Li, L. Wang, H. Yang, Z. Wang, and J. Wang: *Intermetallics (Barking)*, 2019, vol. 114, p. 106592.

74 Y. Chen and Q. Zhou: *Materials Science and Engineering: A*, 2022, vol. 860, p. 144272.

75 M. Zheng, C. Li, L. Zhang, X. Zhang, Z. Ye, X. Yang, and J. Gu: *Materials Science and Engineering: A*, 2022, vol. 840, p. 142933.

76 L. Huang, R. Li, Y. Sun, D. Guan, C. Liang, C. Jiang, J. Chen, D. Wang, and J. Li: *Metals (Basel)*, 2022, vol. 12, p. 1581.

77 M. Zheng, C. Li, X. Zhang, Z. Ye, X. Yang, and J. Gu: *Addit Manuf*, 2021, vol. 37, p. 101660.

78 M.T. Tran, T.H. Nguyen, D.-K. Kim, W. Woo, S.-H. Choi, H.W. Lee, H. Wang, and J.G. Kim: *Materials Science and Engineering: A*, 2021, vol. 828, p. 142110.

79 P. Xue, L. Zhu, P. Xu, H. Lu, S. Wang, Z. Yang, J. Ning, S.L. Sing, and Y. Ren: *J Alloys Compd*, 2022, vol. 928, p. 167169.

80 F. Weng, Y. Chew, Z. Zhu, S. Sui, C. Tan, X. Yao, F.L. Ng, W.K. Ong, and G. Bi: *Compos B Eng*, 2021, vol. 216, p. 108837.

81 Y. Chen, B. Li, B. Chen, and F. Xuan: *Addit Manuf*, 2023, vol. 61, p. 103319.

82 J. Ma, L. Jia, W. Jia, and F. Jin: *Adv Eng Mater*, DOI:10.1002/adem.202200900.

83 K. Zhou, D. Cui, Z. Chai, Y. Zhang, Z. Yang, C. Zhu, Z. Wang, J. Li, and J. Wang: *Addit Manuf*, 2023, vol. 66, p. 103443.

84 P. Niu, R. Li, K. Gan, T. Yuan, S. Xie, and C. Chen: *Metallurgical and Materials Transactions A*, 2021, vol. 52, pp. 753–66.

85 K.-H. Jung, M.T. Tran, Z. Shan, H.W. Lee, S.-K. Hwang, H.G. Kim, and D.-K. Kim: *Journal of Materials Research and Technology*, 2023, vol. 22, pp. 2297–315.

86 H. Sun, Z. Luo, S. Wang, N. Hashimoto, H. Oka, Q. Liu, and S. Isobe: *Mater Lett*, 2023, vol. 347, p. 134644.

87 L. Jinlong, Z. Zhiping, and G. Wenli: *Mater Lett*, 2023, vol. 343, p. 134353.

88 M. Sun, B. Wang, J. Zhang, and B. Lu: *Intermetallics (Barking)*, 2023, vol. 156, p. 107866.

89 R. Li, D. Kong, K. He, and C. Dong: *Scr Mater*, 2023, vol. 230, p. 115401.

90 X. Wang, Z. Ji, R.O. Ritchie, I. Okulov, J. Eckert, and C. Qiu: *Mater Today Adv*, 2023, vol. 18, p. 100371.

91 E.S. Kim, K.R. Ramkumar, G.M. Karthik, S.G. Jeong, S.Y. Ahn, P. Sathiyamoorthi, H. Park, Y.-U. Heo, and H.S. Kim: *J Alloys Compd*, 2023, vol. 942, p. 169062.

92 F. Weng, Y. Chew, Z. Zhu, X. Yao, L. Wang, F.L. Ng, S. Liu, and G. Bi: *Addit Manuf*, 2020, vol. 34, p. 101202.

93 P. Xue, L. Zhu, P. Xu, Y. Ren, B. Xin, S. Wang, Z. Yang, J. Ning, G. Meng, and Z. Liu: *Materials Science and Engineering: A*, 2021, vol. 817, p. 141306.

94 P. Xue, L. Zhu, J. Ning, Y. Ren, Z. Yang, S. Wang, P. Xu, G. Meng, Z. Liu, and B. Xin: *Virtual Phys Prototyp*, 2021, vol. 16, pp. 404–16.

95 X. Bi, R. Li, T. Li, B. Liu, Y. Yuan, P. Zhang, and K. Feng: *Materials Science and Engineering: A*, 2023, vol. 871, p. 144834.

96 P.S. Deshmukh, S. Yadav, G.D. Sathiaraj, and C.P. Paul: *Mater Today Commun*, 2023, vol. 35, p. 106351.

97 Y.S. Kim, H. Chae, W. Woo, D.-K. Kim, D.-H. Lee, S. Harjo, T. Kawasaki, and S.Y. Lee: *Materials Science and Engineering: A*, 2021, vol. 828, p. 142059.

98 S.Y. Ahn, F. Haftlang, E.S. Kim, S.G. Jeong, J.S. Lee, and H.S. Kim: *J Alloys Compd*, 2023, vol. 960, p. 170631.

99 P. ’Gaël, V. ’Alexandre, G. ’Vincent, M. ’Bertrand, T. ’Olivier, G. ’Mathieu, B. ’Peter, P. ’Ron, W. ’Vincent, D. ’Jake, V. ’Alexandre, P. ’David, C. ’Fabian: *Journal of Machine Learning Research*, 2011, vol. 12, pp. 2825–30.

100 L. Ward, A. Agrawal, A. Choudhary, and C. Wolverton: *NPJ Comput Mater*, DOI:10.1038/npjcompumats.2016.28.

101 C. Zhou, Y. Zhang, J. Stasic, Y. Liang, X. Chen, and M. Trtica: *Adv Eng Mater*, DOI:10.1002/adem.202201369.

102 X. Tan, Q. Lu, D. Chen, Z. Wang, H. Chen, X. Peng, H. Xiao, W. Zhang, Z. Liu, L. Guo, and Q. Zhang: DOI:10.20944/preprints202402.0793.v1.

103 A. Hinneburg and D.A. Keim: *Optimal Grid-Clustering: Towards Breaking the Curse of Dimensionality in High-Dimensional Clustering*, 1999.

104 M. Naeem, H. He, F. Zhang, H. Huang, S. Harjo, T. Kawasaki, B. Wang, S. Lan, Z. Wu, F. Wang, Y. Wu, Z. Lu, Z. Zhang, C.T. Liu, and X.-L. Wang: *Cooperative Deformation in High-Entropy Alloys at Ultralow Temperatures*, vol. 6, 2020.

105 E.O. Hall: *Proceedings of the Physical Society. Section B*, 1951, vol. 64, pp. 747–53.

106 J. Ge, C. Chen, R. Zhao, Q. Liu, Y. Long, J. Wang, Z. Ren, and S. Yin: *Mater Des*, 2022, vol. 219, p. 110774.

107 L. Liu, Q. Ding, Y. Zhong, J. Zou, J. Wu, Y.L. Chiu, J. Li, Z. Zhang, Q. Yu, and Z. Shen: *Materials Today*, 2018, vol. 21, pp. 354–61.

108 Y.M. Wang, T. Voisin, J.T. McKeown, J. Ye, N.P. Calta, Z. Li, Z. Zeng, Y. Zhang, W. Chen, T.T. Roehling, R.T. Ott, M.K. Santala, P.J. Depond, M.J. Matthews, A. V. Hamza, and T. Zhu: *Nat Mater*, 2018, vol. 17, pp. 63–70.

109 G.K. Williamson and W.H. Hall: *Acta Metallurgica*, 1953, vol. 1, pp. 22–31.

110 G.K. Williamson and R.E. Smallman: *Philosophical Magazine*, 1956, vol. 1, pp. 34–46.

111 M.A. Meyers and K.K. Chawla: *Mechanical Behavior of Materials*, Cambridge University Press, 2008.

112 E.Orowan: in *Symposium on internal stresses in metals & alloys Institute of Metals*, 1947, pp. 47–59.

113 R. Jarugula, S. Channagiri, S.G.S. Raman, and G. Sundararajan: *Metall Mater Trans A Phys Metall Mater Sci*, 2021, vol. 52, pp. 1901–12.

114 A. Chauhan, F. Bergner, A. Etienne, J. Aktaa, Y. de Carlan, C. Heintze, D. Litvinov, M. Hernandez-Mayoral, E. Oñorbe, B. Radiguet, and A. Ulbricht: *Journal of Nuclear Materials*, 2017, vol. 495, pp. 6–19.

115 N. Kamikawa, K. Sato, G. Miyamoto, M. Murayama, N. Sekido, K. Tsuzaki, and T. Furuhara: *Acta Mater*, 2015, vol. 83, pp. 383–96.

116 S. Queyreau, G. Monnet, and B. Devincre: *Acta Mater*, 2010, vol. 58, pp. 5586–95.

117 J.D. Lord, B. Roebuck, R. Morrell, and T. Lube: *Materials Science and Technology*, 2010, vol. 26, pp. 127–48.

118 B. Yin, S. Yoshida, N. Tsuji, and W.A. Curtin: *Nat Commun*, 2020, vol. 11, p. 2507.

119 C. Zhang, K. Feng, H. Kokawa, B. Han, and Z. Li: *Materials Science and Engineering A*, DOI:10.1016/j.msea.2020.139672.

120 A. Piglione, B. Dovgyy, C. Liu, C.M. Gourlay, P.A. Hooper, and M.S. Pham: *Mater Lett*, 2018, vol. 224, pp. 22–5.

121 K. Lu, A. Chauhan, D. Litvinov, A.S. Tirunilai, J. Freudenberger, A. Kauffmann, M. Heilmaier, and J. Aktaa: *J Mater Sci Technol*, 2022, vol. 100, pp. 237–45.

122 Z. Wang, I. Baker, Z. Cai, S. Chen, J.D. Poplawsky, and W. Guo: *Acta Mater*, 2016, vol. 120, pp. 228–39.

123 C. Varvenne, A. Luque, and W.A. Curtin: *Acta Mater*, 2016, vol. 118, pp. 164–76.

124 L. Xiao, G. Wang, W. Long, P.K. Liaw, and J. Ren: *Eng Fract Mech*, DOI:10.1016/j.engfracmech.2024.109860.

125 Z. Tong, W. Wan, W. Zhou, Y. Ye, J. Jiao, and X. Ren: *Intermetallics (Barking)*, 2022, vol. 151, p. 107710.

# Supplementary Materials

## Experimentally validated and empirically compared machine learning approach for predicting yield strength of additively manufactured multi-principal element alloys from Co-Cr-Fe-Mn-Ni system

Abhinav Chandraker[1], Sampad Barik[1], Nichenametla Jai Sai[2], Ankur Chauhan[1]*

[1]Extreme Environments Materials Group, Department of Materials Engineering, Indian Institute of Science (IISc), Bengaluru, 560012 Karnataka, India

[2]Department of Materials Science and Metallurgy, University of Cambridge, 27 Charles Babbage Road, Cambridge, CB3 0FS, UK

*Corresponding author: ankurchauhan@iisc.ac.in

### 1. Missing value imputation method and data preparation processes

Due to limitations in the available data, some input feature values could not be obtained from the literature. To address this, missing value imputation was performed. When the scan pattern rotation was not reported, a scan pattern rotation of 0° was assumed, indicating no scan rotation. Similarly, if the tensile test direction was unspecified, it was assumed to be along the longitudinal direction (perpendicular to the build direction). For the remaining features, the median values of the respective additive manufacturing routes were calculated and used to replace the missing values. The percentage of missing values in the initial data is presented in Table S1, with the tensile test direction, testing strain rate, and cross-section area notably having a high number of missing values.

Table S1: List of features for which missing values were encountered and their respective statistics.

| Feature | % Missing |
|---|---|
| **Tensile test direction** | 25% |
| **Scan pattern rotation** | 15% |
| **Cross section area** | 12% |
| **Testing strain rate** | 10% |
| **Laser power** | 3% |
| **Scan velocity** | 3% |
| **LPBF VED** | 2% |
| **LMD LED** | 1% |

To maintain consistency within the dataset, a value of zero was assigned to the VED for LPBF-based data points when LMD was used, and vice versa for LED. This approach ensured uniformity in the dataset and accounted for the different energy density measures used in each additive manufacturing technique.

## 2. Analysis of the dataset and its standardization

A box plot was constructed to gain insights into the dataset and compare the variation in parameter ranges for the two printing techniques, as shown in Figure S1. To ensure comparability, all features were standardized using the equation below, which enables a consistent scale for analysis:

$$z = \frac{x-\mu}{\sigma} \qquad \text{Equation 1}$$

Where x is the feature value, μ is the mean feature value, and σ is the standard deviation of the feature.

In the plot, the height of the boxes represents the feature values that lie in the second quartile (Q2) and the third quartile (Q3). This distance is also called the interquartile range (IQR); the middle 50% of the data lies in this range. Visual interpretation of the same is provided in Figure S2.

In the feature plot, standardized mean values above zero for a technique indicate that the mean feature value for that technique was higher compared to the other technique. The plots show that the LPBF printing technique is associated with lower laser power and much higher scan velocity than LMD. Furthermore, the MPEAs produced using LPBF exhibit higher YS, consistent with previous findings [1]. This can be attributed to the significantly higher cooling rates in LPBF, resulting in smaller grain sizes and higher residual stresses in the printed material.

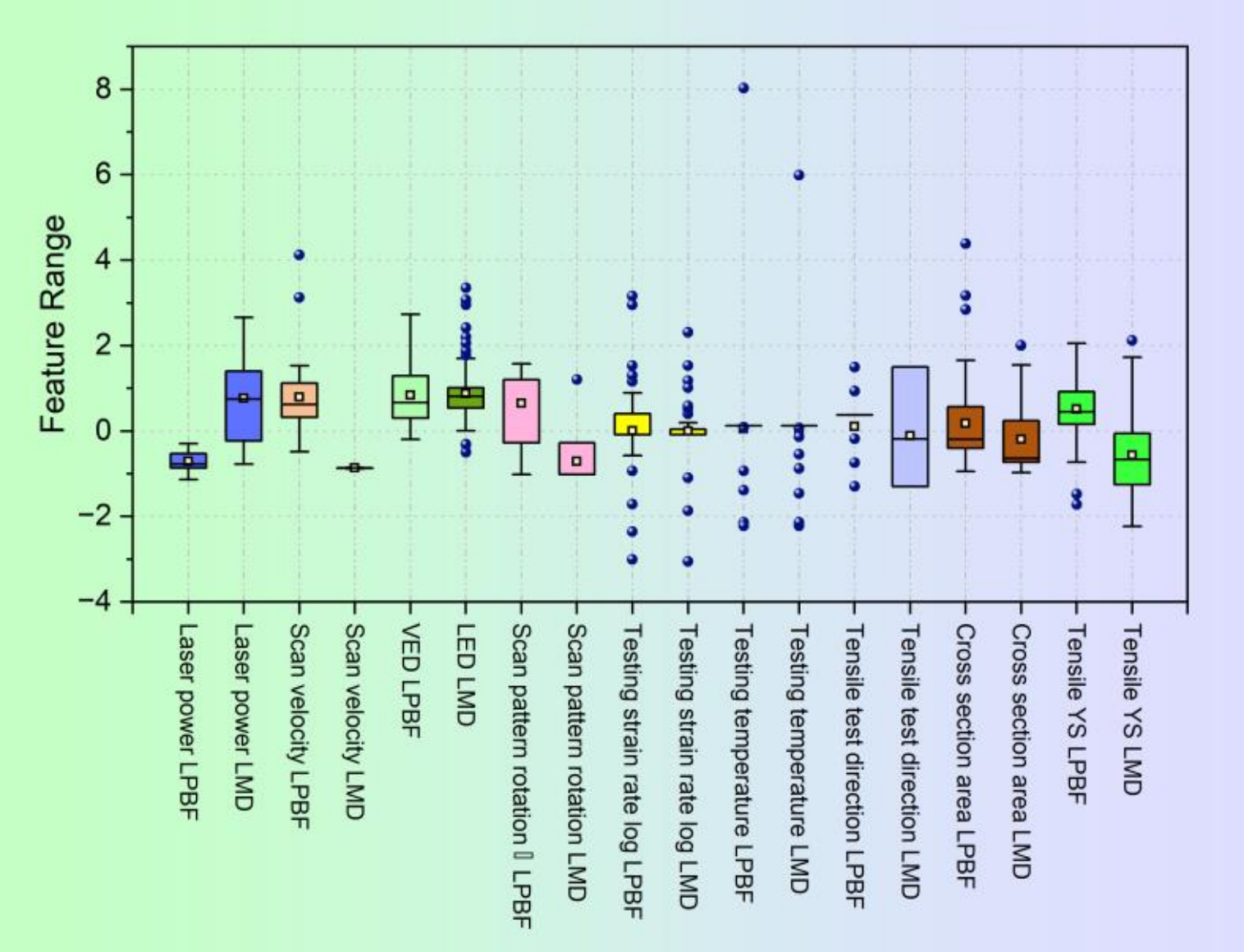

Figure S1: Distribution of standardized feature values for LPBF and LMD routes. The blue sphere represents feature outliers, the horizontal bar represents the median value, and the yellow square represents the mean value. The boxes comprise the middle 50% (Q2 and Q3) of the feature values.

Blue spheres in the feature plots represent feature outliers, identified as data points that lie more than 1.5 times the IQR above Q3 or below Q2. Notably, one experiment [2] in LMD exhibited a very high LED value due to higher laser power and lower scan speed than other works. Similarly, two outliers in the testing temperature feature for both LPBF and LMD plots correspond to tensile testing conducted at 800 °C and 600 °C in Refs. [3][4]. The occurrence of a considerable number of outliers for certain input features suggests that the dataset is sparse, and additional data could enhance the study.

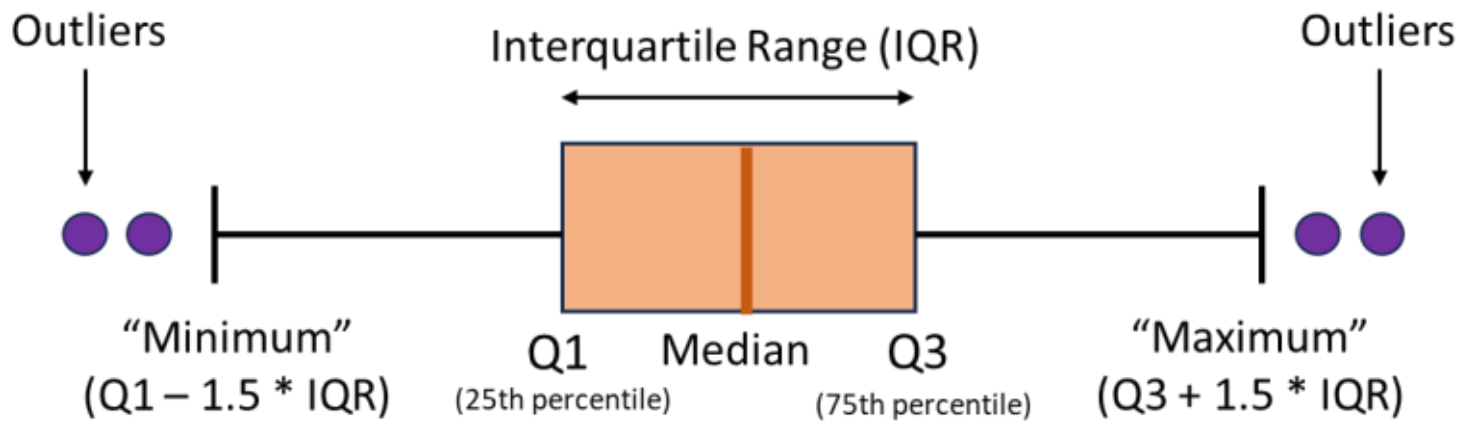

Figure S2: Interpretation of the various components of the box plot.

## 3. Compositional descriptors

Molar average and molar mismatch values [5] of the elemental parameters were considered, as shown in Table S2. Certain composition description functions were used, and their formulas are provided in

Table S3. Descriptors related to enthalpy and entropy were omitted since they primarily pertain to phase prediction, and the alloys in this study are majorly single-phase. The elemental values used for these calculations were taken from references [6,7].

Table S2: Elemental parameters for which mean, and mismatch values were utilized.

| Description | Abbreviation | Description | Abbreviation |
|---|---|---|---|
| Atomic radii | $r$ | p valence electrons | $pVE$ |
| Electronegativity | $\chi$ | d valence electrons | $dVE$ |
| Melting temperature | $T_m$ | Total electrons | $TE$ |
| Column | $c$ | Molar volume | $MV$ |
| 1st Ionization energy | $IE1$ | Density | $\rho$ |
| 2nd Ionization energy | $IE2$ | Atomic number | $AN$ |
| 3rd Ionization energy | $IE3$ | Atomic weight | $AW$ |
| s valence electrons | $sV$ | | |

Table S3: Composition description functions employed and their formulas.

| Description | Abbreviation | Formula |
|---|---|---|
| Local size mismatch | $D.r$ | $\sum_{i=1}^{n} \sum_{j=1,\ i\neq j}^{n} C_i\, C_j * (r_i - r_j)$ |
| Local modulus mismatch | $D.G$ | $\sum_{i=1}^{n} \sum_{j=1,\ i\neq j}^{n} C_i\, C_j * \lvert G_i - G_j \rvert$ |
| Mismatch of local electronegativity | $D.\chi$ | $\sum_{i=1}^{n} \sum_{j=1,\ i\neq j}^{n} C_i\, C_j * \lvert \chi_i - \chi_j \rvert$ |
| Modulus mismatch in strengthening model | $\eta$ | $\sum_{i=1}^{n} \frac{C_i\ * \frac{2\,(G_i - G)}{G_i + G}}{1 + 0.5\ *\ \lvert\, C_i\ * \frac{2\,(G_i - G)}{G_i + G}\ \rvert}$ |
| Energy term in strengthening model | $A$ | $G * \delta r * (1 + \mu) / \,(1 - \mu)$ |
| Peierls Nabarro factor | $F$ | $\frac{2G}{1 - \ \mu}$ |
| Difference in shear modulus | $\delta G$ | $\sqrt{\sum_{i=1}^{n} C_i * (1 - \frac{G_i}{G})^{\,2}}$ |
| Lattice distortion energy | $LDE$ | $\frac{1}{2}\ E\ *\ \delta r$ |
| Valence electron concentration | $VEC$ | $\sum_{i=1}^{n} C_i * VEC_i$ |
| Intra lattice elasticity difference | $\gamma$ | $\left(1 - \sqrt{\frac{(r + r_{min})^2 - \ r^2}{(r + r_{min})^2}}\right) \Bigg/ \left(1 - \sqrt{\frac{(r + r_{max})^2 - \ r^2}{(r + r_{max})^2}}\right)$ |

**Notes:** $C_i, r_i, G_i$ and $VEC_i$ represent mole ratio, atomic radius, shear modulus, and valence electron concentration, respectively. $r, \chi, G, E,$ and $\mu$ represent the molar average value of the atomic radius, electronegativity, shear modulus, Young's modulus, and Poisson's ratio for the specific composition. $r_{min}$ and $r_{max}$ are the minimum atomic radius and maximum atomic radius of the elements, respectively.

The molar average value of a parameter for a composition was calculated using -

$$\bar{X} = \ \textstyle\sum_{i=1}^{n} C_i\, X_i \qquad \text{Equation 2}$$

The molar mismatch between the element parameter values for a composition was calculated using the following relation [5].

$$\delta_X \ = \ \sqrt{\textstyle\sum_{i=1}^{n} C_i * (1 - \frac{X_i}{\bar{X}})^{\,2}} \qquad \text{Equation 3}$$

## 4. Encoding technique

Categorical features were manually encoded with numerical values. The encoded values corresponding to each category are provided in Table S4.

Table S4: Numerical encoding employed for the values of the categorical features

| Categorical feature | Values |
|---|---|
| **Scan pattern rotation** | 0° – 0; 180° – 1; 90° – 2; 90°-checkboard – 3; 45°-checkboard – 4; 60° – 5; 67° – 6; 67°-checkboard - 7 |
| **Tensile test direction** | X without scan rotation – 0; 45° to X and Y – 1; Y without scan rotation – 2; X with scan rotation - 3; 45° to X and Z – 4; Z – 5 |

## 5. Input feature selection for dimensionality reduction

Feature selection was conducted in two steps. First, the most highly correlated features were removed to reduce redundancy, using a threshold value of 0.9. In this step, the more correlated feature from each highly correlated pair was sequentially removed [8]. This process reduced the initial 54 features to 31, with 23 features being removed. Next, an RF algorithm was used to fit the data and rank the features by importance.

To investigate the effect of the number of features on model performance, starting with the highest-ranked six features, the most important features were added sequentially, and the models' accuracies were recorded. It was observed that amongst the four models employed, RF consistently showed the highest accuracy; its model performance is shown in Figure S3 for the number of features varying from 6 to 20.

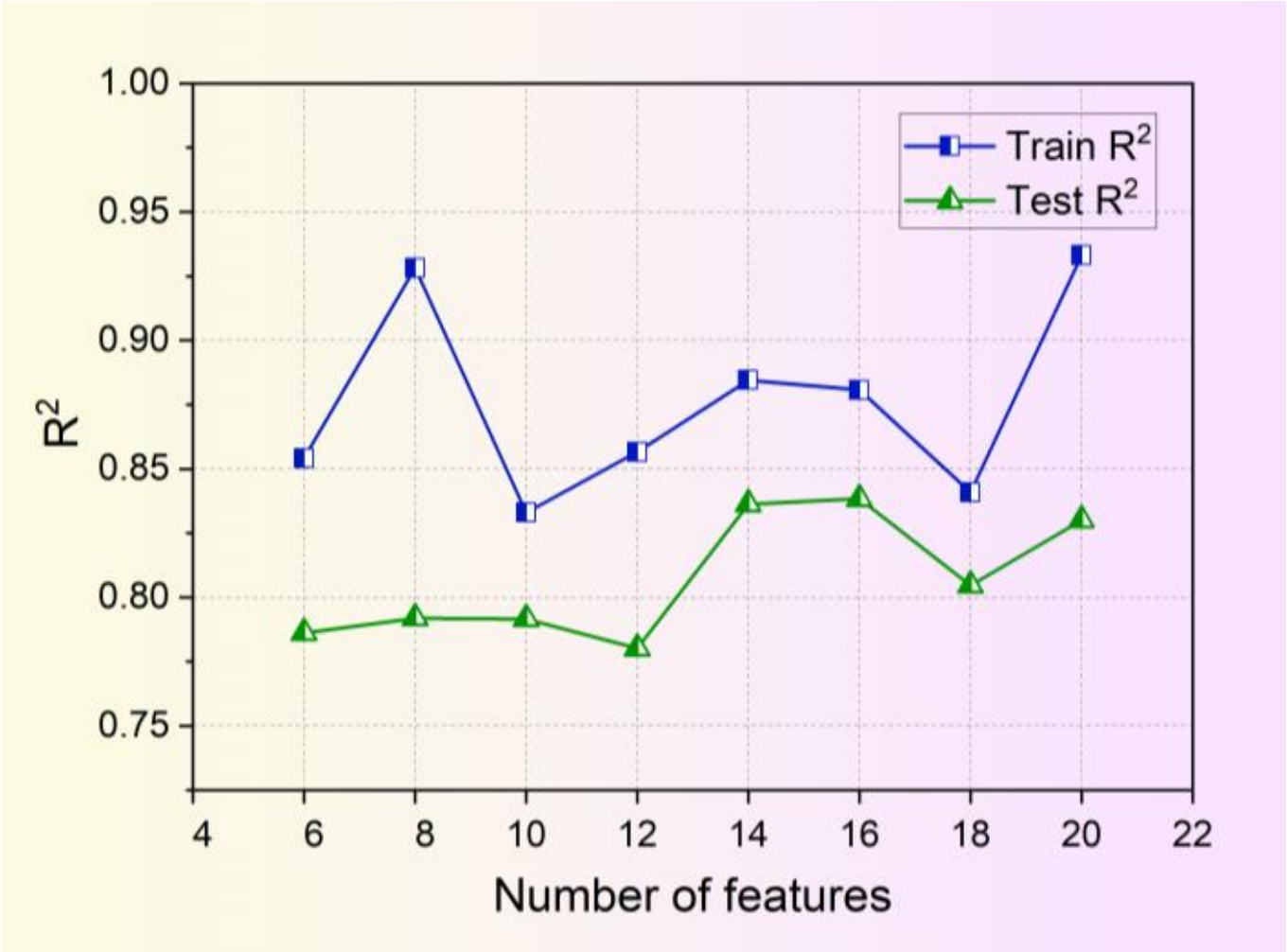

Figure S3: Effect of the number of features employed on testing accuracy for the RF model, which consistently showed higher accuracy than the other models.

The highest accuracy was observed with 16 features. The $R^2$ decreased from 0.84 to 0.78 when reducing the features from 14 to 12 by removing Peierl's Nabarro stress and testing strain rate. The importance and probable effects of these features on YS are discussed in detail in the main manuscript, **'Input feature importance' section**.

All the processing and testing-related parameters were retained in the list of 16 features, while only the composition descriptors were excluded. The low ranking of these compositional descriptors is attributed to the significantly larger effect of other parameters on the prediction. These parameters are also discussed in the main manuscript, **'Input feature importance' section**.

### 5.1. Correlation matrix

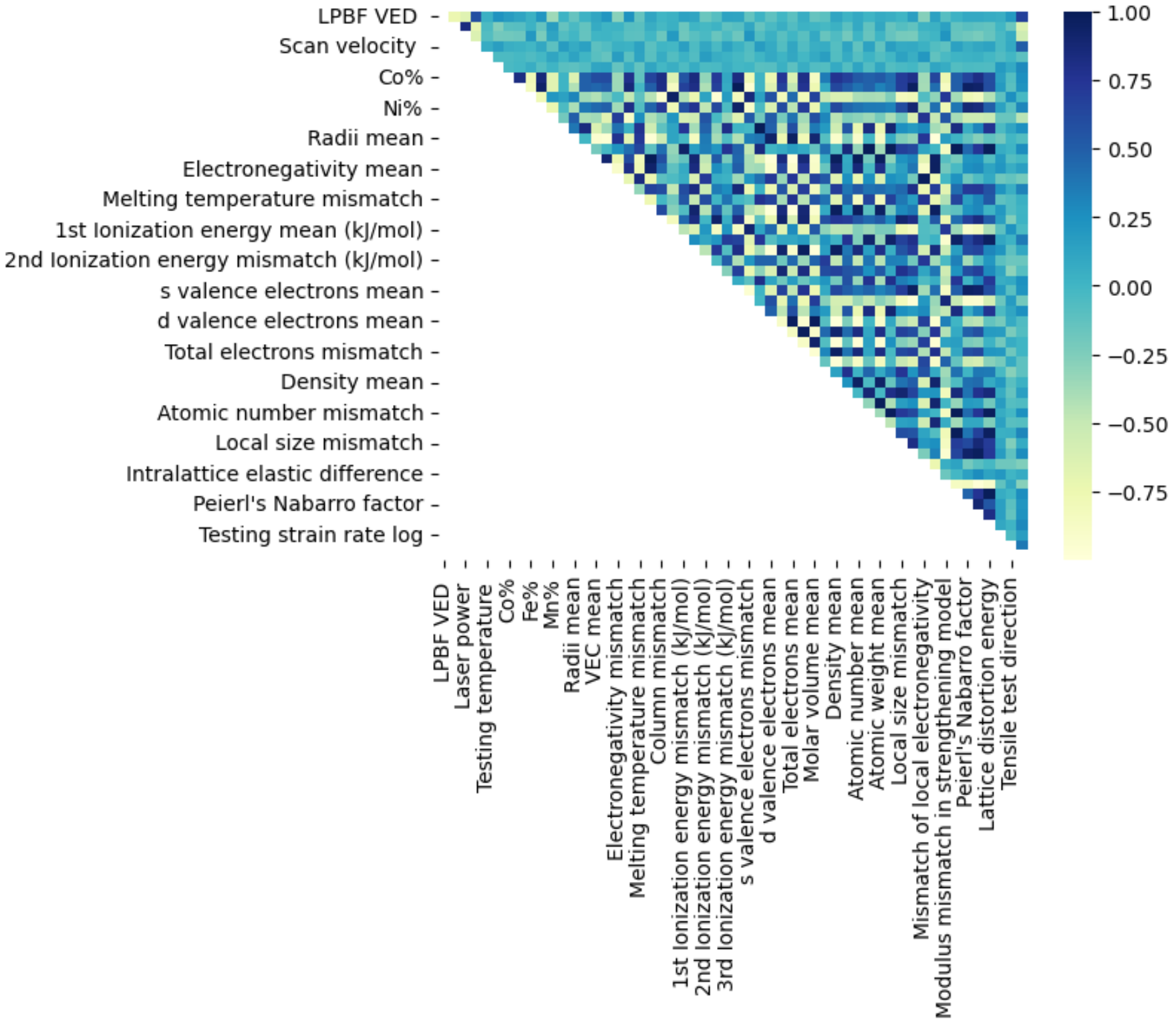

Figure S4: Visualization of correlation coefficients of all possible feature pairs in the initial dataset.

## 6. Brief descriptions of the employed ML algorithms

### 6.1. Random Forest (RF)

Random Forest (RF) is an ensemble algorithm that utilizes multiple decision trees (DT) to make predictions. It works on the divide-and-conquer approach [9]. Its main benefit is overcoming the overfitting tendency observed when using a single DT. Each RF DT consists of decision and leaf nodes [10]. Decision nodes have decision-making constraints; the outcomes are stored in the leaf nodes. In RF, the training dataset for each tree is created by sampling the training set with replacement, a technique known as bootstrapping. This means that each DT is trained on a different subset of the dataset. For regression problems, the final output is determined by averaging the outputs of all the trees. This averaging process helps mitigate overfitting that may occur in some of the individual trees, thus reducing the overall impact of overfitting. Several hyperparameters are important in RF, including the maximum depth of the trees, the maximum number of features considered for splitting at each decision node, and the number of trees in the ensemble. These hyperparameters play a crucial role in controlling the complexity and performance of the RF model.

### 6.2. Gradient Boosting (GBoost)

Gradient Boosting (GBoost) represents an alternative form of ensembled algorithm based on decision trees (DT), and offers a distinct approach compared to the RF algorithm. In GBoost, each DT aims to sequentially correct the errors made by its predecessor, resulting in a different tree structure than RF [10]. Unlike RF, the DT in GBoost is relatively shallow. The GBoost algorithm begins by calculating the average of the output feature values, which serves as the initial predicted output for all data points. It then evaluates the discrepancy between the predicted and actual output values and constructs a DT to fit these errors. The predicted output is subsequently updated as the sum of the initial average value and the error calculated by the first DT. This process is iteratively repeated for each subsequent DT. As a result, the algorithm generates a series of DTs sequentially, incorporating errors from previous predictions and the actual outputs. Important hyperparameters in GBoost include the maximum depth of the trees, the learning rate, and the number of trees. These hyperparameters play a crucial role in controlling the complexity and performance of the GBoost model. The overall output is computed by aggregating the predictions of all the DTs generated in the iterative process as follows:

*The net output = average of actual output (first predicted output) +*

$$\sum_{i=1}^{n} learning\ rate \times error\ from\ i^{th}\ DT$$

### 6.2 Extreme Gradient Boosting (XGBoost)

Extreme Gradient Boosting (XGBoost) is an advancement from the GBoost algorithm, built on a similar underlying principle. However, in XGBoost, the trees employed are not conventional DT but are specifically designed as XGBoost trees. These trees introduce the "amount of say" concept to calculate the gain at each decision node. The amount of say at each node is calculated using the following equation:

$$\text{Amount of say at each node} = \frac{(sum\ of\ residuals)^2}{number\ of\ residuals/errors + \lambda}$$

The gain in XGBoost is determined using the following formula:

$$\text{Gain} = \sum(amount\ of\ say\ of\ i^{th}\ daughter\ node) - \text{the amount of say of parent decision node}$$

Consequently, the XGBoost trees utilize the gain concept to select constraints at each decision node, considering their potential for higher gain. The amount of say plays a crucial role in determining the gain and influences the constraints chosen. Several hyperparameters control the behavior of XGBoost, including the number of trees, eta (learning rate), lambda (regularization parameter), gamma (parameter for pruning trees), and maximum depth.

### 6.3. Support Vector Regression (SVR)

Support Vector Regression (SVR) is a supervised machine learning algorithm used for regression tasks; it essentially extends the concepts of Support Vector Machine (SVM) to regression problems. It aims to find a hyperplane that best fits the data within a certain margin or threshold of error.

The linear hypothesis is represented as: $h(x)= wTx+b$

Where $h(x)$ is the predicted output, $w$ is the weight vector, $x$ is the input feature vector, $b$ is the bias term.

A loss function is used to measure the error between the predicted and actual values, with the objective of minimizing the loss function while maximizing the margin.

The most commonly used loss function is the epsilon-insensitive loss function, given by:

$L\epsilon\ (y, h(x)) = \max\ (0, |y-h(x)| - \epsilon)$

Where $y$ is the true output, $h(x)$ is the predicted output, $\epsilon$ is the tolerance margin.

It uses a kernel trick to handle nonlinear relationships between input features and the target variable. By mapping the input features into a higher-dimensional space using a kernel function, SVR can find a hyperplane that better fits the data even when the relationship is nonlinear. Some common kernel functions are linear, polynomial, and radial basis function (RBF), and their choice depends on the nature of the data and the problem at hand. It also has a regularization parameter and is particularly useful when dealing with datasets with complex relationships or outliers.

### 6.4. SHapley Additive ExPlanations (SHAP)

SHapley Additive ExPlanations (SHAP) is a technique rooted in game theory that provides insights into the correlation between each input feature and the output feature [11]. It offers a means to determine the importance and the associated relationship of each input feature with the output feature. Additionally, SHAP allows for examining the impact of each input feature on every instance or observation.

For each instance, a SHAP value is assigned to each input feature, where larger SHAP values indicate a stronger influence on predicting the output feature. To calculate the SHAP value for a target input feature, subsets of input features that include the target feature are considered, and the ML model is applied to these subsets to predict the output. The target input feature is then removed from the subsets, and the ML model is retrained and used for predictions. The final SHAP value for the target input feature of a specific instance is determined as the weighted average of the differences in predicted output between the subsets of input features with and without the target feature across all subsets. Mathematically, the SHAP value can be represented as follows:

$$\Phi_i(f, x) = \sum_{z' \subseteq x} \frac{(|z' - 1|!\,(M - |z'|)!}{M!}[f_x(z') - f_x(z'/i)]$$

Where

$\Phi_i(f, x)$ = SHAP value of a target input feature 'i' by considering a model 'f' for a particular instance 'x'

$z'$= subset of the input features having a target input feature

$f_x(z')$ = output of the model with a subset of input features having a target input feature

$f_x(z'/i)$ = output of the model with a subset of input features without having a target input feature

## 7. Hyperparameter tuning method – Bayesian Search

Hyperparameter tuning involves searching for the optimal set of hyperparameters that effectively models a given dataset. Techniques such as Grid search and Randomized search explore different sets of hyperparameters individually and without prior information. In contrast, Bayesian search leverages past search evaluations by storing and utilizing that information. It constructs a probabilistic model based on it, often called a surrogate function, which is then optimized. The hyperparameter set that performs best on the surrogate function is evaluated, and the evaluation result is used to update and enhance the surrogate function. This iterative process continues, allowing for a more efficient and effective search for the optimal hyperparameters. Bayesian search offers a faster and more robust approach to finding the optimum hyperparameters.

## 8. List of tuned hyper-parameters used for employed ML algorithms

Table S5: Hyper-parameters used for the RF algorithm.

| **Hyper-parameters** | **Values** |
|---|---|
| *Max depth* | 7 |
| *Max features* | 1 |
| *Min_samples_leaf* | 1 |
| *Min_ samples_split* | 2 |
| *N_estimators* | 806 |
| *OOB scores* | False |

Table S6: Hyper-parameters used for the GBoost algorithm.

| **Hyper-parameters** | **Values** |
|---|---|
| *Learning rate* | 0.025 |
| *Max depth* | 1 |
| *Min_samples_leaf* | 1 |
| *N_estimators* | 518 |

Table S7: Hyper-parameters used for the XGBoost algorithm.

| **Hyper-Parameters** | **Values** |
|---|---|
| *Colsample_bynode* | 0.68 |
| *Colsample_bytree* | 0.1 |
| *Eta* | 0.1 |
| *Gamma* | 0.0 |
| *Lambda* | 1.0 |
| *Max depth* | 4 |
| *Min_child_weight* | 5 |
| *N_estimators* | 260 |
| *Subsample* | 0.4 |

Table S8: Hyper-parameters used for the SVR algorithm.

| **Hyper-Parameters** | **Values** |
|---|---|
| *C* | 1.78 |
| *Epsilon* | 0.01 |
| *Gamma* | 0.01 |
| *Kernel* | rbf |

## 9. Model robustness investigation using multiple data splits

Investigating the impact of data selection on the performance of the models is crucial. To reliably evaluate the models' sensitivity to data splits, 25 additional sets of training and testing datasets were created with the same 4:1 splitting ratio as the original split. The hyperparameters optimized for the initial split were used to train and test the two best-performing models, RF and XGBoost, on these new splits. The resulting $R^2$ and RMSE values are shown in Figure S5a and Figure S5b, respectively.

A low sensitivity to data splitting is desirable as it indicates that the model is robust and consistently performs well even when the data changes. The training accuracies for the new splits were comparable to those of the initial split, indicating successful model training. The testing accuracies exhibited slight variations, demonstrating marginal sensitivity to data splitting. The mean $R^2$ for the new splits was 0.71 for RF and 0.73 for XGBoost, with standard deviations of 0.09 and 0.08, respectively. The mean RMSE was 96 MPa for RF and 92 MPa for XGBoost, with 19.2 and 16.8 MPa standard deviations, respectively.

Notably, while the testing accuracy of XGBoost was lower than that of RF for the original split for which hyperparameters were tuned, its chosen hyperparameters were less sensitive to the data split employed. This underscores the importance of assessing model robustness alongside evaluating model performance.

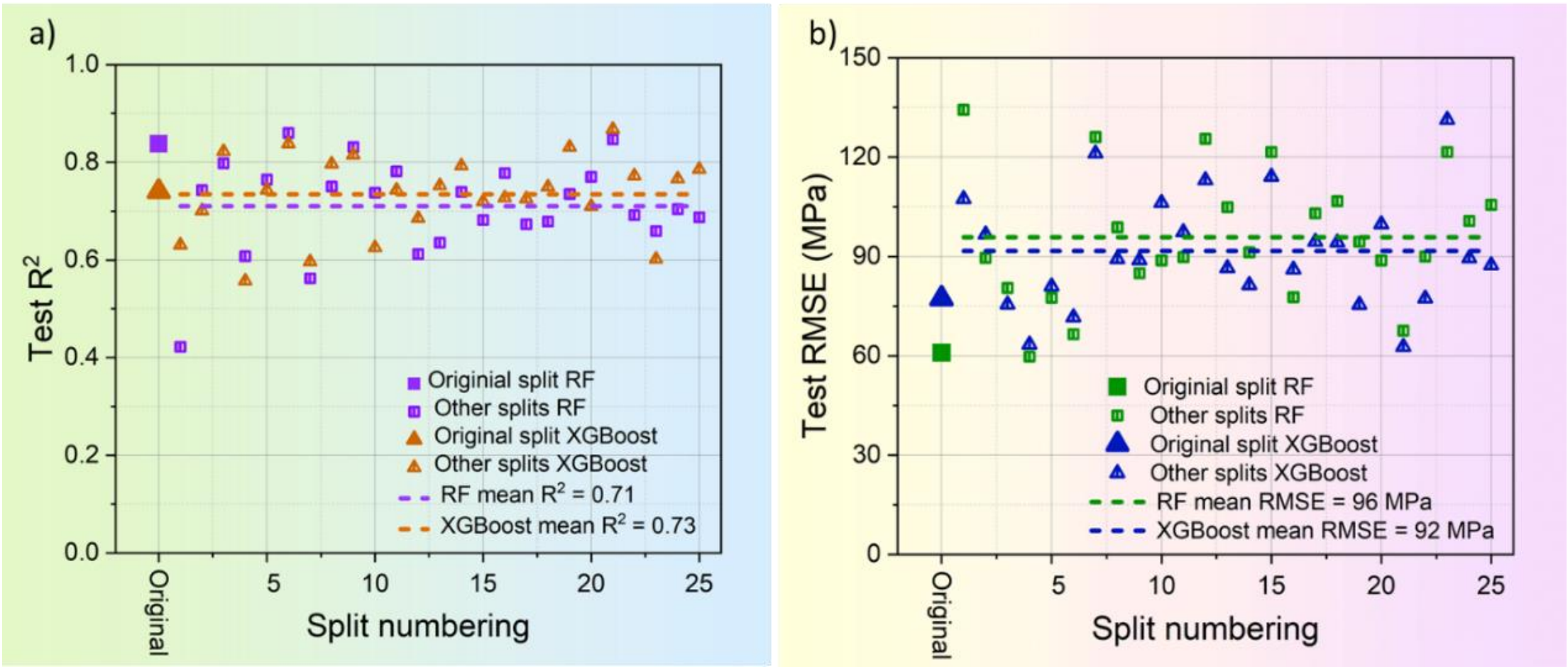

Figure S5: Data split sensitivity of model hyperparameters considering a) testing R2 and b) testing RMSE for RF and XGBoost algorithms.

## 10. Microstructural details from the literature and empirical strengthening contributions

Table S9: Microstructural and material-specific constants details from the literature utilized for estimating strengthening contributions (few values were unreported and have been mentioned as NA (not available).

| MPEA (Additive manufacturing route) | $k$ (MPa µm$^{1/2}$) | $d$ (µm) | $\alpha$ | $M$ | $G$ (GPa) | $b$ (nm) | $\rho$ (*$10^{13}$ m$^{-2}$) | Reference |
|---|---|---|---|---|---|---|---|---|
| $Co_{20}Cr_{20}Fe_{20}Mn_{20}Ni_{20}$ (LPBF) | 494 | 42.9 | 0.2 | 3.06 | 81 | 0.255 | 3.2 | [12] |
| $Co_{20}Cr_{20}Fe_{20}Mn_{20}Ni_{20}$ (LPBF) | 494 | 12.9 | 0.2 | 3.06 | 81 | 0.255 | 21.6 | [13] |
| $Co_{20}Cr_{20}Fe_{20}Mn_{20}Ni_{20}$ (LPBF) | 494 | 14.4 | 0.2 | 3.06 | 81 | 0.255 | 28.4 | [14] |
| $Co_{20}Cr_{20}Fe_{20}Mn_{20}Ni_{20}$ (LPBF) | 494 | 23.2 | 0.2 | 3.06 | 81 | 0.255 | 28.4 | [14] |
| $Co_{20}Cr_{20}Fe_{20}Mn_{20}Ni_{20}$ (LPBF) | 494 | 21.1 | 0.2 | 3.06 | 81 | 0.255 | 28.4 | [14] |
| $Co_{20}Cr_{20}Fe_{20}Mn_{20}Ni_{20}$ (LPBF) | 494 | 15 | 0.2 | 3.01 | 81 | 0.255 | 148 | [15] |
| $Co_{20}Cr_{20}Fe_{20}Mn_{20}Ni_{20}$-1%C (LPBF) | 494 | 30.5 | 0.2 | 3.06 | 81 | 0.255 | 85 | [16] |
| $Co_{20}Cr_{20}Fe_{20}Mn_{20}Ni_{20}$-1%C (LPBF) | 494 | 20.3 | 0.2 | 3.06 | 81 | 0.255 | 58 | [16] |
| $Co_{33.3}Cr_{33.3}Ni_{33.3}$ (LPBF) | 265 | 0.5 | 0.2 | 3.06 | 90 | 0.252 | 7.5 | [17] |
| $Co_{33.3}Cr_{33.3}Ni_{33.3}$ (LPBF) | 265 | 0.6 | 0.2 | 3.06 | 90 | 0.252 | 12.6 | [17] |
| $Co_{33.3}Cr_{33.3}Ni_{33.3}$ (LPBF) | 265 | 0.7 | 0.2 | 3.06 | 90 | 0.252 | 13.2 | [17] |
| $Co_{33.3}Cr_{33.3}Ni_{33.3}$ (LPBF) | 265 | 25.6 | 0.2 | 3.06 | 90 | 0.252 | 54 | [18] |
| $Fe_{60}Co_{15}Ni_{15}Cr_{10}$ (LPBF) | 339 | 12 | 0.2 | 2.85 | NA | NA | 42.6 | [18] |
| $Co_{20}Cr_{20}Fe_{20}Mn_{20}Ni_{20}$ (LMD) | 494 | 41 | 0.2 | 3.19 | 81 | 0.255 | 120 | [15] |
| $Co_{20}Cr_{20}Fe_{20}Mn_{20}Ni_{20}$ (LMD) | 494 | NA | 0.2 | 3.06 | 81 | 0.255 | 10.1 | [19] |
| $Co_{20}Cr_{20}Fe_{20}Mn_{20}Ni_{20}$ (LMD) | 494 | NA | 0.2 | 3.06 | 81 | 0.255 | 10.1 | [19] |
| $Co_{20}Cr_{20}Fe_{20}Mn_{20}Ni_{20}$ (LMD) | 494 | NA | 0.2 | 3.06 | 81 | 0.255 | 1 | [20] |
| $Co_{20}Cr_{20}Fe_{20}Mn_{20}Ni_{20}$ (LMD) | 494 | 35 | 0.2 | 3.2 | 81 | 0.255 | 5 | [21] |
| $Co_{20}Cr_{20}Fe_{20}Mn_{20}Ni_{20}$ (LMD) | 494 | 45.8 | 0.2 | 3.35 | 81 | 0.255 | 4.1 | [21] |
| $Co_{20}Cr_{20}Fe_{20}Mn_{20}Ni_{20}$ (LMD) | 494 | 70.1 | 0.2 | 3.21 | 81 | 0.255 | 60.1 | [22] |
| $Co_{33.3}Cr_{33.3}Ni_{33.3}$ (LMD) | 265 | 161 | 0.2 | 3.05 | 90 | 0.252 | 120 | [23] |
| $Co_{33.3}Cr_{33.3}Ni_{33.3}$ (LMD) | 265 | 93.9 | 0.2 | 3.06 | 90 | 0.252 | 4 | [24] |
| $Co_{33.3}Cr_{33.3}Ni_{33.3}$ (LMD) | 265 | 5.3 | 0.2 | 3.06 | 90 | 0.252 | 35.8 | [25] |
| $Co_{33.3}Cr_{33.3}Ni_{33.3}$ (LMD) | 265 | 4.4 | 0.2 | 3.06 | 90 | 0.252 | 17.9 | [25] |
| $Co_{33.3}Cr_{33.3}Ni_{33.3}$ (LMD) | 265 | 5.3 | 0.2 | 3.06 | 90 | 0.252 | 35.8 | [26] |
| $Co_{33.3}Cr_{33.3}Ni_{33.3}$ (LMD) | 265 | 8.3 | 0.2 | 2.15 | 90 | 0.252 | 27.8 | [27] |
| $Co_{33.3}Cr_{33.3}Ni_{33.3}$ (LMD) | 265 | 16.9 | 0.2 | 2.15 | 90 | 0.252 | 27.8 | [27] |
| $Co_{33.3}Cr_{33.3}Ni_{33.3}$ (LMD) | 265 | 14.9 | 0.2 | 2.15 | 90 | 0.252 | 28 | [28] |
| $Co_{33.3}Cr_{33.3}Ni_{33.3}$ (LMD) | 265 | 9.2 | 0.2 | 2.15 | 90 | 0.252 | 34.3 | [28] |
| $Co_{10}Cr_{10}Fe_{49.5}Mn_{30}C_{0.5}$ (LMD) | 573 | 72.4 | 0.2 | 3.06 | NA | NA | 10 | [29] |

Table S10: Contributions calculated from various strengthening mechanisms for the empirically predicted data points.

| MPEA (Additive manufacturing route) | Lattice Friction (MPa) | Hall-Petch strengthening (MPa) | Dislocation strengthening (MPa) | Orowan strengthening (MPa) |
|---|---|---|---|---|
| $Co_{20}Cr_{20}Fe_{20}Mn_{20}Ni_{20}$ (LPBF) | 125 | 75 | 71 | 263 |
| $Co_{20}Cr_{20}Fe_{20}Mn_{20}Ni_{20}$ (LPBF) | 125 | 138 | 186 | 0 |
| $Co_{20}Cr_{20}Fe_{20}Mn_{20}Ni_{20}$ (LPBF) | 125 | 130 | 213 | 0 |
| $Co_{20}Cr_{20}Fe_{20}Mn_{20}Ni_{20}$ (LPBF) | 125 | 103 | 213 | 0 |
| $Co_{20}Cr_{20}Fe_{20}Mn_{20}Ni_{20}$ (LPBF) | 125 | 108 | 213 | 0 |
| $Co_{20}Cr_{20}Fe_{20}Mn_{20}Ni_{20}$ (LPBF) | 125 | 128 | 478 | 0 |
| $Co_{20}Cr_{20}Fe_{20}Mn_{20}Ni_{20}$-1%C (LPBF) | 203 | 89 | 369 | 201 |
| $Co_{20}Cr_{20}Fe_{20}Mn_{20}Ni_{20}$-1%C (LPBF) | 203 | 110 | 304 | 161 |
| $Co_{33.3}Cr_{33.3}Ni_{33.3}$ (LPBF) | 218 | 382 | 120 | 0 |
| $Co_{33.3}Cr_{33.3}Ni_{33.3}$ (LPBF) | 218 | 335 | 156 | 0 |
| $Co_{33.3}Cr_{33.3}Ni_{33.3}$ (LPBF) | 218 | 323 | 160 | 0 |
| $Co_{33.3}Cr_{33.3}Ni_{33.3}$ (LPBF) | 218 | 52 | 323 | 0 |
| $Fe_{60}Co_{15}Ni_{15}Cr_{10}$ (LPBF) | 103 | 98 | 270 | 0 |
| $Co_{20}Cr_{20}Fe_{20}Mn_{20}Ni_{20}$ (LMD) | 125 | 77 | 456 | 0 |
| $Co_{20}Cr_{20}Fe_{20}Mn_{20}Ni_{20}$ (LMD) | 125 | 113 | 127 | 0 |
| $Co_{20}Cr_{20}Fe_{20}Mn_{20}Ni_{20}$ (LMD) | 125 | 93 | 127 | 0 |
| $Co_{20}Cr_{20}Fe_{20}Mn_{20}Ni_{20}$ (LMD) | 125 | 42 | 40 | 0 |
| $Co_{20}Cr_{20}Fe_{20}Mn_{20}Ni_{20}$ (LMD) | 125 | 83 | 94 | 0 |
| $Co_{20}Cr_{20}Fe_{20}Mn_{20}Ni_{20}$ | 125 | 73 | 88 | 0 |
| (LMD) | | | | |
| $Co_{20}Cr_{20}Fe_{20}Mn_{20}Ni_{20}$ (LMD) | 125 | 59 | 325 | 0 |
| $Co_{33.3}Cr_{33.3}Ni_{33.3}$ (LMD) | 218 | 21 | 479 | 0 |
| $Co_{33.3}Cr_{33.3}Ni_{33.3}$ (LMD) | 218 | 27 | 88 | 0 |
| $Co_{33.3}Cr_{33.3}Ni_{33.3}$ (LMD) | 218 | 115 | 263 | 22 |
| $Co_{33.3}Cr_{33.3}Ni_{33.3}$ (LMD) | 218 | 127 | 186 | 34 |
| $Co_{33.3}Cr_{33.3}Ni_{33.3}$ (LMD) | 218 | 116 | 263 | 0 |
| $Co_{33.3}Cr_{33.3}Ni_{33.3}$ (LMD) | 218 | 92 | 163 | 0 |
| $Co_{33.3}Cr_{33.3}Ni_{33.3}$ (LMD) | 218 | 65 | 163 | 0 |
| $Co_{33.3}Cr_{33.3}Ni_{33.3}$ (LMD) | 218 | 69 | 163 | 0 |
| $Co_{33.3}Cr_{33.3}Ni_{33.3}$ (LMD) | 218 | 87 | 181 | 0 |
| $Co_{10}Cr_{10}Fe_{49.5}Mn_{30}C_{0.5}$ (LMD) | 179 | 67 | 196 | 138 |

Table S11: Comparison of the experimental YS with that predicted using empirical linear, RMS-1, RMS summations, and best-performing ML-based RF model.

| MPEA (Additive manufacturing route) | Experimental YS (MPa) | Linear summation (MPa) | RMS-1 summation (MPa) | RMS summation (MPa) | RF model predicted YS (MPa) |
|---|---|---|---|---|---|
| $Co_{20}Cr_{20}Fe_{20}Mn_{20}Ni_{20}$ (LPBF) | 624 | 534 | 473 | 309 | 533 |
| $Co_{20}Cr_{20}Fe_{20}Mn_{20}Ni_{20}$ (LPBF) | 510 | 448 | 448 | 263 | 542 |
| $Co_{20}Cr_{20}Fe_{20}Mn_{20}Ni_{20}$ (LPBF) | 550 | 468 | 468 | 279 | 599 |
| $Co_{20}Cr_{20}Fe_{20}Mn_{20}Ni_{20}$ (LPBF) | 479 | 441 | 441 | 267 | 533 |
| $Co_{20}Cr_{20}Fe_{20}Mn_{20}Ni_{20}$ (LPBF) | 443 | 446 | 446 | 269 | 483 |
| $Co_{20}Cr_{20}Fe_{20}Mn_{20}Ni_{20}$ (LPBF) | 624 | 731 | 731 | 511 | 544 |
| $Co_{20}Cr_{20}Fe_{20}Mn_{20}Ni_{20}$-1%C (LPBF) | 829 | 862 | 712 | 475 | 543 |
| $Co_{20}Cr_{20}Fe_{20}Mn_{20}Ni_{20}$-1%C (LPBF) | 741 | 778 | 657 | 415 | 518 |
| $Co_{33.3}Cr_{33.3}Ni_{33.3}$ (LPBF) | 689 | 721 | 721 | 456 | 616 |
| $Co_{33.3}Cr_{33.3}Ni_{33.3}$ (LPBF) | 695 | 709 | 709 | 429 | 618 |
| $Co_{33.3}Cr_{33.3}Ni_{33.3}$ (LPBF) | 700 | 700 | 700 | 421 | 613 |
| $Co_{33.3}Cr_{33.3}Ni_{33.3}$ (LPBF) | 630 | 593 | 593 | 393 | 634 |
| $Fe_{60}Co_{15}Ni_{15}Cr_{10}$ (LPBF) | 456 | 471 | 471 | 305 | 637 |
| $Co_{20}Cr_{20}Fe_{20}Mn_{20}Ni_{20}$ (LMD) | 573 | 659 | 659 | 480 | 555 |
| $Co_{20}Cr_{20}Fe_{20}Mn_{20}Ni_{20}$ (LMD) | 387 | 365 | 365 | 211 | 407 |
| $Co_{20}Cr_{20}Fe_{20}Mn_{20}Ni_{20}$ (LMD) | 351 | 345 | 345 | 201 | 402 |
| $Co_{20}Cr_{20}Fe_{20}Mn_{20}Ni_{20}$ (LMD) | 290 | 207 | 207 | 138 | 665 |
| $Co_{20}Cr_{20}Fe_{20}Mn_{20}Ni_{20}$ (LMD) | 175 | 302 | 302 | 177 | 395 |
| $Co_{20}Cr_{20}Fe_{20}Mn_{20}Ni_{20}$ (LMD) | 174 | 286 | 286 | 169 | 225 |
| $Co_{20}Cr_{20}Fe_{20}Mn_{20}Ni_{20}$ (LMD) | 485 | 509 | 509 | 353 | 232 |
| $Co_{33.3}Cr_{33.3}Ni_{33.3}$ (LMD) | 557 | 718 | 718 | 527 | 584 |
| $Co_{33.3}Cr_{33.3}Ni_{33.3}$ (LMD) | 368 | 333 | 333 | 237 | 407 |
| $Co_{33.3}Cr_{33.3}Ni_{33.3}$ (LMD) | 621 | 618 | 597 | 361 | 563 |
| $Co_{33.3}Cr_{33.3}Ni_{33.3}$ (LMD) | 579 | 565 | 534 | 315 | 551 |
| $Co_{33.3}Cr_{33.3}Ni_{33.3}$ (LMD) | 621 | 596 | 596 | 360 | 519 |
| | 347 | 473 | 473 | 287 | 453 |
| $Co_{33.3}Cr_{33.3}Ni_{33.3}$ (LMD) | | | | | |
| $Co_{33.3}Cr_{33.3}Ni_{33.3}$ (LMD) | 361 | 445 | 445 | 280 | 443 |
| $Co_{33.3}Cr_{33.3}Ni_{33.3}$ (LMD) | 400 | 450 | 450 | 281 | 526 |
| $Co_{33.3}Cr_{33.3}Ni_{33.3}$ (LMD) | 380 | 486 | 486 | 296 | 402 |
| $Co_{10}Cr_{10}Fe_{49.5}Mn_{30}C_{0.5}$ (LMD) | 645 | 580 | 486 | 307 | 404 |

## References

[1] Y. Liu, J. Ren, S. Guan, C. Li, Y. Zhang, S. Muskeri, Z. Liu, D. Yu, Y. Chen, K. An, Y. Cao, W. Liu, Y. Zhu, W. Chen, S. Mukherjee, T. Zhu, W. Chen, Microstructure and mechanical behavior of additively manufactured CoCrFeMnNi high-entropy alloys: Laser directed energy deposition versus powder bed fusion, Acta Mater 250 (2023). https://doi.org/10.1016/j.actamat.2023.118884.

[2] Z. Qiu, C. Yao, K. Feng, Z. Li, P.K. Chu, Cryogenic deformation mechanism of CrMnFeCoNi high-entropy alloy fabricated by laser additive manufacturing process, International Journal of Lightweight Materials and Manufacture 1 (2018) 33–39. https://doi.org/10.1016/j.ijlmm.2018.02.001.

[3] R. Li, D. Kong, K. He, C. Dong, Superior thermal stability and strength of additively manufactured CoCrFeMnNi high-entropy alloy via NbC decorated sub-micro dislocation cells, Scr Mater 230 (2023) 115401. https://doi.org/10.1016/j.scriptamat.2023.115401.

[4] J. Li, S. Xiang, H. Luan, A. Amar, X. Liu, S. Lu, Y. Zeng, G. Le, X. Wang, F. Qu, C. Jiang, G. Yang, Additive manufacturing of high-strength CrMnFeCoNi high-entropy alloys-based composites with WC addition, J Mater Sci Technol 35 (2019) 2430–2434. https://doi.org/10.1016/j.jmst.2019.05.062.

[5] X. Tan, Q. Lu, D. Chen, Z. Wang, H. Chen, X. Peng, H. Xiao, W. Zhang, Z. Liu, L. Guo, Q. Zhang, Machine Learning Prediction of Phase and Tensile Properties of High Entropy Alloys Manufactured by Selective Laser Melting, (2024). https://doi.org/10.20944/preprints202402.0793.v1.

[6] Winter Mark, WebElements, (n.d.). https://www.webelements.com/ (accessed June 17, 2024).

[7] Gray Theodore, periodictable.com, (n.d.). https://periodictable.com/ (accessed June 17, 2024).

[8] C. Yang, C. Ren, Y. Jia, G. Wang, M. Li, W. Lu, A machine learning-based alloy design system to facilitate the rational design of high entropy alloys with enhanced hardness, Acta Mater 222 (2022). https://doi.org/10.1016/j.actamat.2021.117431.

[9] S.R. Sukhdeve, Step Up for Leadership in Enterprise Data Science and Artificial Intelligence with Big Data: Illustrations with R and Python, Independently Published, 2020.

[10] S.G. A.C. Muller, Introduction to Machine Learning with Python: A Guide for Data Scientists, O'Reilly Media, 2016.

[11] P.G. A. Gramegna, SHAP and LIME: An Evaluation of Discriminative Power in Credit Risk, Front Artif Intell, 2021.

[12] P. Chen, C. Yang, S. Li, M.M. Attallah, M. Yan, In-situ alloyed, oxide-dispersion-strengthened CoCrFeMnNi high entropy alloy fabricated via laser powder bed fusion, Mater Des 194 (2020) 108966. https://doi.org/10.1016/j.matdes.2020.108966.

[13] Z.G. Zhu, Q.B. Nguyen, F.L. Ng, X.H. An, X.Z. Liao, P.K. Liaw, S.M.L. Nai, J. Wei, Hierarchical microstructure and strengthening mechanisms of a CoCrFeNiMn high entropy alloy additively manufactured by selective laser melting, Scr Mater 154 (2018) 20–24. https://doi.org/10.1016/j.scriptamat.2018.05.015.

[14] C. Zhang, K. Feng, H. Kokawa, Z. Li, Correlation between microstructural heterogeneity and anisotropy of mechanical properties of laser powder bed fused CoCrFeMnNi high entropy alloy, Materials Science and Engineering: A 855 (2022) 143920. https://doi.org/10.1016/j.msea.2022.143920.

[15] E.S. Kim, K.R. Ramkumar, G.M. Karthik, S.G. Jeong, S.Y. Ahn, P. Sathiyamoorthi, H. Park, Y.-U. Heo, H.S. Kim, Cryogenic tensile behavior of laser additive manufactured CoCrFeMnNi high entropy alloys, J Alloys Compd 942 (2023) 169062. https://doi.org/10.1016/j.jallcom.2023.169062.

[16] J.M. Park, J. Choe, J.G. Kim, J.W. Bae, J. Moon, S. Yang, K.T. Kim, J.H. Yu, H.S. Kim, Superior tensile properties of 1%C-CoCrFeMnNi high-entropy alloy additively manufactured by selective laser melting, Mater Res Lett 8 (2020) 1–7. https://doi.org/10.1080/21663831.2019.1638844.

[17] J. Ge, C. Chen, R. Zhao, Q. Liu, Y. Long, J. Wang, Z. Ren, S. Yin, Strength-ductility synergy of CoCrNi medium-entropy alloy processed with laser powder bed fusion, Mater Des 219 (2022) 110774. https://doi.org/10.1016/j.matdes.2022.110774.

[18] P. Niu, R. Li, K. Gan, T. Yuan, S. Xie, C. Chen, Microstructure, Properties, and Metallurgical Defects of an Equimolar CoCrNi Medium Entropy Alloy Additively Manufactured by Selective Laser Melting, Metallurgical and Materials Transactions A 52 (2021) 753–766. https://doi.org/10.1007/s11661-020-06121-4.

[19] Y.L. Wang, L. Zhao, D. Wan, S. Guan, K.C. Chan, Additive manufacturing of TiB2-containing CoCrFeMnNi high-entropy alloy matrix composites with high density and enhanced mechanical properties, Materials Science and Engineering: A 825 (2021) 141871. https://doi.org/10.1016/j.msea.2021.141871.

[20] H.G. Li, Y.J. Huang, W.J. Zhao, T. Chen, J.F. Sun, D.Q. Wei, Q. Du, Y.C. Zou, Y.Z. Lu, P. Zhu, X. Lu, A.H.W. Ngan, Overcoming the strength-ductility trade-off in an additively manufactured CoCrFeMnNi high entropy alloy via deep cryogenic treatment, Addit Manuf 50 (2022) 102546. https://doi.org/10.1016/j.addma.2021.102546.

[21] M. Zheng, C. Li, X. Zhang, Z. Ye, X. Yang, J. Gu, The influence of columnar to equiaxed transition on deformation behavior of FeCoCrNiMn high entropy alloy fabricated by laser-based directed

energy deposition, Addit Manuf 37 (2021) 101660. https://doi.org/10.1016/j.addma.2020.101660.

[22] S.Y. Ahn, F. Haftlang, E.S. Kim, S.G. Jeong, J.S. Lee, H.S. Kim, Boost in mechanical strength of additive manufactured CoCrFeMnNi HEA by reinforcement inclusion of B4C nano-particles, J Alloys Compd 960 (2023) 170631. https://doi.org/10.1016/j.jallcom.2023.170631.

[23] M.T. Tran, T.H. Nguyen, D.-K. Kim, W. Woo, S.-H. Choi, H.W. Lee, H. Wang, J.G. Kim, Effect of hot isostatic pressing on the cryogenic mechanical properties of CrCoNi medium entropy alloy processed by direct energy deposition, Materials Science and Engineering: A 828 (2021) 142110. https://doi.org/10.1016/j.msea.2021.142110.

[24] P. Xue, L. Zhu, P. Xu, H. Lu, S. Wang, Z. Yang, J. Ning, S.L. Sing, Y. Ren, Microstructure evolution and enhanced mechanical properties of additively manufactured CrCoNi medium-entropy alloy composites, J Alloys Compd 928 (2022) 167169. https://doi.org/10.1016/j.jallcom.2022.167169.

[25] F. Weng, Y. Chew, Z. Zhu, S. Sui, C. Tan, X. Yao, F.L. Ng, W.K. Ong, G. Bi, Influence of oxides on the cryogenic tensile properties of the laser aided additive manufactured CoCrNi medium entropy alloy, Compos B Eng 216 (2021) 108837. https://doi.org/10.1016/j.compositesb.2021.108837.

[26] F. Weng, Y. Chew, Z. Zhu, X. Yao, L. Wang, F.L. Ng, S. Liu, G. Bi, Excellent combination of strength and ductility of CoCrNi medium entropy alloy fabricated by laser aided additive manufacturing, Addit Manuf 34 (2020) 101202. https://doi.org/10.1016/j.addma.2020.101202.

[27] P. Xue, L. Zhu, P. Xu, Y. Ren, B. Xin, S. Wang, Z. Yang, J. Ning, G. Meng, Z. Liu, CrCoNi medium-entropy alloy thin-walled parts manufactured by laser metal deposition: Microstructure evolution and mechanical anisotropy, Materials Science and Engineering: A 817 (2021) 141306. https://doi.org/10.1016/j.msea.2021.141306.

[28] P.S. Deshmukh, S. Yadav, G.D. Sathiaraj, C.P. Paul, Nano to macro-mechanical properties of laser directed energy deposited CoCrNi medium entropy alloy, Mater Today Commun 35 (2023) 106351. https://doi.org/10.1016/j.mtcomm.2023.106351.

[29] Y. Chew, Z.G. Zhu, F. Weng, S.B. Gao, F.L. Ng, B.Y. Lee, G.J. Bi, Microstructure and mechanical behavior of laser aided additive manufactured low carbon interstitial Fe49.5Mn30Co10Cr10C0.5 multicomponent alloy, J Mater Sci Technol 77 (2021) 38–46. https://doi.org/10.1016/j.jmst.2020.11.026.